\documentclass[a4paper,12pt]{article}

    \usepackage{indentfirst}
    \usepackage{booktabs}
    \usepackage{color}
    \usepackage{graphicx}
    \usepackage[hang,small,tight]{subfigure}
    \usepackage{amsmath}
    \usepackage{amssymb}
    \usepackage{mathrsfs}
    \usepackage{rotating}
    \usepackage{multirow}
    \usepackage{html}
    \usepackage{url}
    \usepackage{harvard}
    \usepackage[ansinew]{inputenc}

    \newcommand{\bs}{\boldsymbol}

    \newcommand{\reais}{\mathbb{R}}
    
  \title{Accounting for spatially varying directional effects \\ in spatial covariance structures}

  \author{Joaquim Henriques Vianna Neto \\ Universidade Federal de Juiz de Fora, Brazil
\\Alexandra M. Schmidt\footnote{Addresses for correspondence: A. M. Schmidt is Professor in  Statistics of the Deparment of Statistical Methods, of the Federal University of Rio de Janeiro, Brazil. E-mail: {\tt  alex@im.ufrj.br} and {\tt http://www.dme.ufrj.br/$\sim$alex}. \newline J. H. Vianna Neto, Universidade Federal de Juiz de Fora (UFJF), Juiz de Fora, Minas Gerais, Brazil. E-mail: {\tt joaquim.neto@ufjf.edu.br} and {\tt http://www.ufjf.br/joaquim\_neto/}. \newline
   P. Guttorp is Professor in Statistics at the University of Washington, USA, and Guest Professor at the Norwegian Computing Center, Norway. E-mail: {\tt peter@stat.washington.edu} and {\tt http://www.stat.washington.edu/peter/}.} \\ Universidade Federal do Rio de Janeiro,  Brazil
\\ Peter Guttorp \\ University of Washington, USA and \\ Norwegian Computing Center, Oslo, Norway}

  \date{}

\begin{document}

  \maketitle

\vspace{-0.7cm}

  \begin{abstract}
   Wind direction plays an important role in the spread of pollutant levels over a geographical region. We discuss how to include wind directional information in the covariance function of spatial models. We follow the spatial convolution approach initially proposed by  Higdon and co-authors, wherein a spatial process is described by a convolution between a smoothing kernel and a white noise process. We propose two different ways of accounting for wind direction in the kernel function. For comparison purposes, we also consider a more flexible kernel parametrization, that makes use of latent processes which vary smoothly across the region. Inference procedure follows the Bayesian paradigm, and uncertainty about parameter estimation is naturally accounted for when performing spatial interpolation. We analyze ozone levels observed at a monitoring network in the Northeast of the USA. Samples from the posterior distribution under our proposed models are obtained much faster when compared to the kernel based on latent processes. Our models provide better results, in terms of model fitting and spatial interpolation, when compared to simple isotropic and geometrical anisotropic models. Despite the small number of parameters, our proposed models provide fits which are comparable to those obtained under the kernel based on latent processes. 
\end{abstract}





\section{Introduction}

\subsection{Motivation \label{sec:motivation}}

As described in the homepage of the United States Environmental Protection Agency (EPA) ({\tt http://www.epa.gov/airnow/airaware/day2-detail.html}) ``large weather systems dictate the predominant wind direction, which can have a considerable effect on the quality of air in a specific city or region. Air quality can worsen in a particular region  if the wind is blowing from a region that contains numerous sources of pollution. If the winds are coming from areas with little or no pollution, they can make your air quality better. Very light winds or no wind, such as those in a strong high pressure system, can be a problem for urban areas, because all the pollution that a city creates stays in one place."

Usually, air pollution data are observed at fixed points (monitoring stations) of a region of interest. It is common practice to model these data using spatial models (see e.g. \citeasnoun**{BANERJEE2004}), which allow for  spatial correlation among observations at different locations. 
More specifically, in spatial statistics, one usually assumes that observations are partial
realizations of a stochastic process $\{Z(s), s \in G\}$, with $G \in \reais^C$, where commonly $C=2$, and the components of the location vector $s$ are geographical coordinates. Frequently, it is assumed
that $Z(\cdot)$ follows a Gaussian process (GP) with a stationary and isotropic
covariance structure; that is the spatial distribution is unchanged under rotation about the origin, translation of the origin of the index set. 

When modelling air pollutant data it is expected from the quote above that wind direction has a local effect in the process of interest and the assumption of stationarity and isotropy seem unreaslitic. Moreover, in the spatial setting, the usual aim is to make
spatial interpolation to unobserved locations of interest, based on
observed values at monitored locations. This interpolation is
heavily based on the specification of the mean and covariance
structures of the GP. In environmental problems the assumption of
stationary covariance structures is commonly violated due to local
influences in  the covariance structure of the process.

In this paper, we aim at proposing non-stationary spatial models that account for wind direction in the covariance structure of air pollutant data. Winds are vector quantities and can be split into orthogonal components. Usually, wind direction is reported through  two horizontal components, the $u$ component which represents the east-west direction, and the $v$ component which represents the north-south direction.  
The challenge is how to account for this directional information when common correlation functions are usually functions of the norm of a vector. It is expected that if the wind at  two arbitrary locations is blowing in the same direction, the process at the two locations should be more highly correlated than the process at locations which have winds blowing in different directions. We propose two different ways of accounting for wind direction in the covariance function of a spatial process.

\begin{figure}[!htb]
\begin{center}
\subfigure{\includegraphics[scale=.7]{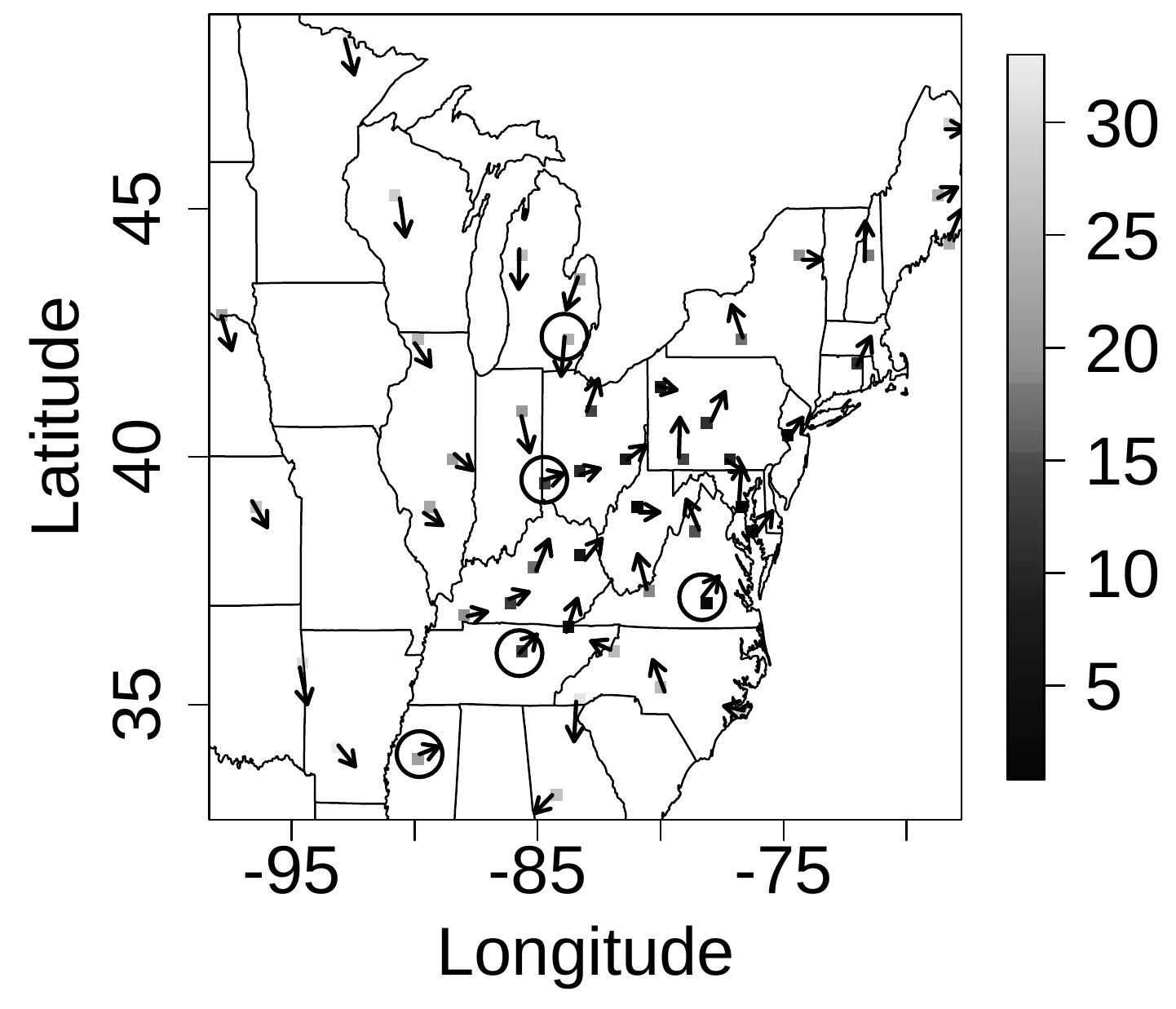}} 
\end{center}
\caption{Observed values of ozone (solid squares in grayscale), and respective wind direction (arrows) for December, 11th, 2008, 3pm. For predictive purposes, circled locations are left out from the inference procedure performed in Section \ref{sec:dataana}. \label{fig:data}}
\end{figure}

In order to show the ability of our proposed models in capturing the influence of wind direction on the covariance structure of air pollutant data, we analyze levels of ozone observed at a particular time of a day, at monitoring locations located in the northeast of the USA. As described in the {\tt AIRNow} website ({\tt http://www.airnow.gov}), ozone forms near ground level when air pollutants (emitted by sources such as cars, power plants, and chemical plants) react chemically in the presence of sunlight. 
EPA maintains the {\em Clean Air Status and Trends Network} (CASTNET) which is ``a national air quality monitoring network designed to provide data to assess trends in air quality, atmospheric deposition, and ecological effects due to changes in air pollutant emissions". The dataset we analyze here was downloaded from the site \newline {\tt http://java.epa.gov/castnet/epa\_jsp/prepackageddata.jsp/\#ozone}. We consider  $n=48$ monitoring stations, all of them containing information about ozone and wind measurements. Figure \ref{fig:data} shows the locations of the  monitoring sites and the observed data.

There are many different alternatives available in the literature to model flexible spatial covariance structures. In the last decade many of them have focused on the   fact that a Gaussian process can be described as a convolution between a white noise process and a kernel function. Next Subsection reviews these approaches in more detail, as our proposal in Section \ref{sec:proposta} is based on the  spatial convolution approach.

\subsection{A brief review of spatial convolution models \label{sec:convol}}

A stochastic process can be constructed by convolving a white noise process $W(\cdot )$ with a smoothing kernel $k(\cdot )$ by defining the process as $Y(s)=\int_G k(h)W(h)dh$.  \citeasnoun**{HIGDON1999} 
 propose non-stationary spatial models based on this framework by allowing the kernel to vary smoothly with location, that is, they propose
\begin{equation}\label{eqn:conv1}
  Y(s) = \int\limits_{G } {k_s\left( h \right) W \left( h \right)dh}.
\end{equation}
 The kernel $k_s(.)$ is an arbitrary function;  it is common practice to define it based on density functions or nonnegative functions that decay monotonically from their mode \cite{PACIOREK2006}. \citeasnoun**{PACIOREK2006} show that when $k_s(.)$ is  a Gaussian kernel centered at $s$, and with covariance matrix $\Sigma(s)$, the resultant nonstationary covariance structure of the process $Y(.)$ is equal to
 $$
 Cov(Y(s_i),Y(s_j))=\sigma^2 |\Sigma(s_i)|^{1/4}|\Sigma(s_j)|^{1/4}\left|  \frac{\Sigma(s_i)+\Sigma(s_j)}{2}\right|^{-1/2} \exp(Q_{ij}), \mbox{ for } s_i, s_j \in G,
 $$
where $Q_{ij}=(s_i-s_j)^T\left(\frac{\Sigma(s_i)+\Sigma(s_j)}{2}   \right)^{-1} (s_i-s_j)$. \citeasnoun**{HIGDON1999} propose to model the covariance matrix $\Sigma(s_i)$ through the connection between a bivariate normal distribution and its one standard deviation ellipse, such that
\begin{equation}\label{eqn:ups}
\Sigma \left( s \right) =  R\left( s \right)^T \left[ {\begin{array}{*{20}c}
   {\frac{{\sqrt {4\varepsilon ^2  + \left\| \psi(s)  \right\|^4 \pi ^2 } }}{{2\pi }} + \frac{{\left\| \psi(s)  \right\|^2 }}{2}} & 0  \\
   0 & {\frac{{\sqrt {4\varepsilon ^2  + \left\| \psi(s)  \right\|^4 \pi ^2 } }}{{2\pi }} - \frac{{\left\| \psi(s)  \right\|^2 }}{2}}  \\
\end{array}} \right]R\left( s \right),
\end{equation}
where $\left\| {\psi (s)} \right\|=\psi_1(s)^2+\psi_2(s)^2$,  and the rotation matrices, $R(s)$, are given by
\begin{equation*}
R\left( s \right) = \left[ {\begin{array}{*{20}c}
   {\cos \left( {\omega \left( s \right)} \right)} & {\sin \left( {\omega \left( s \right)} \right)}  \\
   { - \sin \left( {\omega \left( s \right)} \right)} & {\cos \left( {\omega \left( s \right)} \right)}  \\
\end{array}} \right], 
\end{equation*}
with $\omega \left( s \right) = \arctan \left( {\psi _2 \left( s \right)/\psi _1 \left( s \right)} \right)$. 
\citeasnoun**{HIGDON1999} let $\psi(s)=(\psi_1(s),\psi_2(s))'$ be the focal points, such that $-\psi(s)=(-\psi_1(s),-\psi_2(s))$, $\forall s\in G$, define a set of ellipses centered at the origin, all of them with a fixed area $\varepsilon$, which is common to all locations. They suggest to fix this area after some exploratory data analysis. 

Because this resultant covariance structure is infinitely differentiable, \citeasnoun**{PACIOREK2006} generalize the kernel convolution approach by introducing a new class of nonstationary covariance functions. This class includes a nonstationary version of the Mat\'ern covariance function, which is of the form
\begin{eqnarray}
C(s_i,s_j)=\sigma^2 \frac{1}{\Gamma(\nu)2^{\nu-1}}|\Sigma(s_i)|^{-1/4}|\Sigma(s_j)|^{-1/4} \left|\frac{\Sigma(s_i)+\Sigma(s_j)}{2}  \right|^{-1/2}\left(2\sqrt{\nu Q_{ij}}\right)^\nu \kappa_{\nu}\left(2\sqrt{\nu Q_{ij}}\right), \label{eq:covarMatern}
\end{eqnarray}
with $Q_{ij}$ as above, $\nu>0$ is the shape parameter, and $\kappa_{\nu}(.)$ is the modified Bessel function of the second kind of order $\nu$. 
Again the main issue is how to model $\Sigma(s_i)$.
\citeasnoun**{PACIOREK2006} model $\Sigma(s_i)$ using the spectral decomposition, by defining $\Sigma(s_i)=\Gamma_i\Lambda_i\Gamma_i^T$, where $\Lambda_i$ is a diagonal matrix of eigenvalues, $\lambda_1(s_i)$ and $\lambda_2(s_i)$, and $\Gamma_i$ is an eigenvector matrix.  They impose some restriction on the elements of $\Gamma_i$ and $\Lambda_i$ to limit the number of hyperparameters and improve mixing of the chains.
Similar to \citeasnoun**{HIGDON1999}, they make use of latent spatial processes to allow $\Gamma_i$ and $\Lambda_i$ to vary smoothly across the spatial region.

Note that when modelling $\Sigma(s_i)$ following either \citeasnoun**{HIGDON1999}, or \citeasnoun**{PACIOREK2006}, the number of parameters to be estimated increases linearly  with the number of observed locations. This is because the covariance matrix $\Sigma(s_i)$ modelled either as in equation (\ref{eqn:ups}), or as in (\ref{eq:covarMatern}), is built through  latent Gaussian processes leading to a great number of parameters to be estimated.

The advantage of specifying a Gaussian kernel in equation (\ref{eqn:conv1}), or the nonstationary version of the Mat\'ern covariance function, is that the resultant covariance function has a closed form. If a different kernel function is used, and it does not result in a closed form of the covariance function, one can approximate the continuous model in (\ref{eqn:conv1}) through a finite sum by defining
\begin{equation}\label{eqn:discretized1}
Y(s) = \sum\limits_{l = 1}^m {k_s \left(h_l \right)W \left( {h_l } \right)},
\end{equation}
where $W(\cdot )$ is a white noise process defined over a fixed grid of $m$ points,  $\{h_1,\cdots,h_m\}$, overlaid on $G$. See \citeasnoun**{HIGDON1998} for a spatio-temporal application. As pointed out by \citeasnoun**{HIGDON2002}, discretized convolutions are economical parameterizations and can greatly facilitate computation, since a small number of processes, $W(h_1),\cdots,W(h_m)$, effectively control the entire spatial process $Y(\cdot )$, even though $Y(\cdot )$ may be required at any location in $G$. Besides being used to approximate the integral and to alleviate the computation burden associated with model fitting, the discretized convolution framework given in (\ref{eqn:discretized1}) has been extended to allow $W(\cdot )$ to be a spatial dependent process itself. Examples of such approaches are discussed, for example, in \citeasnoun**{FUENTES2002}, \citeasnoun**{LEE2005} and \citeasnoun**{CRESSIE2006}.

Usually, the above models are estimated under the Bayesian framework, and Markov chain Monte Carlo (MCMC) methods are used to obtain samples from the resultant posterior distribution. Even under the discretized version of the convolution structure these models have a  great number of parameters to be estimated. All of the parameters involved in the covariance structure of the process $Y(.)$ do not result on a known full conditional posterior distribution. Usually Metropolis-Hastings  steps are used to obtain samples from such full conditionals, and it is quite challenging to tune the variance of the proposal distribution in order to obtain reasonable acceptance rates and reach convergence of the chains. 
Indeed, the models above require fixing some hyperparameters which are problem specific. And from our experience, the algorithms seem quite sensitive to the choice of these hyperparameters.
See Section 5.1 of \citeasnoun**{PACIOREK2006} for further discussion about the computational demands of such models.

\subsection{Literature review on including covariates in spatial covariance functions}

Sources of nonstationarity might be related to local influences of  some covariate effects in the covariance structure of the underlying spatial process. Therefore, it seems reasonable to investigate how to include covariate information in the covariance structure of spatial processes.

\citeasnoun**{verhoef:peterson:theobald:2006} propose spatial models whose covariance structures incorporate flow and stream distance through the use of spatial moving averages.  \citeasnoun**{cooley:nychka:naveau:2007} use covariates (but not geographic coordinates) to model extreme precipitation.

\citeasnoun**{CALDER2008} makes use of wind measurements in the kernel convolution approach of \citeasnoun**{HIGDON1998}. However, she only uses the information of wind direction at a single location to fix the covariance matrix of the Gaussian kernels of the convolution.

\citeasnoun**{schmidt:guttorp:ohagan:2011} extend the work
of \citeasnoun**{schmidt:ohagan:2003} 
by allowing a latent space, wherein isotropy holds,
to be of dimension greater than 2; more specifically, geographic coordinates and covariates are used to define the axis of the latent space. \citeasnoun**{schmidt:guttorp:ohagan:2011} also propose a simpler version of the higher dimensional latent space model,  by defining a projection onto the $\reais^2$ manifold which makes use of covariate information in the covariance structure of the process.
\citeasnoun**{schmidt:rodriguez:2011} use this projection approach to describe the spatial covariance structure when modelling multivariate counts of fish species observed across the shores of a lake.

\citeasnoun**{reichetal:2011}  make use of the convolution approach proposed by \citeasnoun**{FUENTES2002} to introduce local covariate information in the covariance structure of spatio-temporal processes. However, they do not discuss how to include a  directional covariate in their model.

The main aim of this paper is to extend the convolution approach revised in Section \ref{sec:convol} to allow the kernel function in equation (\ref{eqn:conv1}) to depend on the direction the wind is blowing at a particular location. This significantly reduces the number of parameters to be estimated, while still allowing for a flexible covariance structure. 
 We propose two different ways of capturing the directional behaviour of wind in the covariance structure of spatial processes.

This paper is organized as follows. In Section \ref{sec:proposta}, we propose two different ways of accounting for wind direction in the covariance structure of spatial processes induced by a convolution approach. Then in Section \ref{sec:alternative}, following \citeasnoun**{PACIOREK2006} we  propose an alternative model for $\Sigma(s)$ in equation (\ref{eq:covarMatern}).  The aim is to compare a more flexible model with those proposed  in Sections \ref{sec:Gaussian}  and \ref{sec:kernelsdiscretized}. 
Inference procedure is based on the Bayesian paradigm and this is described in Section \ref{sec:inference}. Therein, spatial interpolation to unmonitored locations of interest, and model comparison are also discussed.
In Section \ref{sec:dataana}, we analyze the ozone data presented in Section \ref{sec:motivation}. Therein we compare the performance of the proposed models of Section \ref{sec:proposta} with standard isotropic and  geometric anisotropic models. Finally, Section \ref{sec:conclusao} concludes.

\section{Proposed Models}\label{sec:proposta}

Assume that observations are a partial realization of a random
process $\{Z(s), s \in G \}$, with $G \in \reais^p$, where usually
$p=1,\, 2,$ or $3$. More specifically, let
\begin{equation}\label{eqn:geral}
  Z(s)=\mu(s)+Y(s)+\epsilon(s), 
\end{equation}
where $\mu(s)$ represents the mean structure of the process and
usually is a function of location $s$,  $\epsilon(s)$ is a white noise process, such  that  $\epsilon(s) \sim N(0,\tau^2)$, $\forall \, s$, and is usually present to capture measurement error.
 We assume $Y(s)$ is independent of $\epsilon(s)$, $\forall s$, and $Y(\cdot )$ represents a latent spatial process which captures any spatial structure left after adjusting for the effect of possible covariates included in $\mu(\cdot )$.  We assume $Y(\cdot )$ is a Gaussian process described by
\begin{equation} \label{eqn-proccontinuo}
  Y(s) = \int\limits_{G} {k_{s,x}(h)W(h)dh}, \text{ for } s,h \in G \subset \reais^2,
\end{equation}
where $k_{s,x}(h)$ is a spatially varying kernel function that depends not only on the location $s$ but also on the observed  covariate $x(s)$. 

We now propose two different formulations for $k_{s,x}(.)$. We consider $x(s)=(u(s),v(s))'$, where $u(s)$ and $v(s)$ are, respectively, the $u$ and $v$ components of the wind direction at location $s$. We start by presenting an alternative model for $\Sigma(s)$ in equation (\ref{eq:covarMatern}). Indeed, for our model, we denote this matrix as  $\Sigma(s,x)$ as it also depends on the wind direction.   Then we introduce a more flexible kernel function which also accounts for  wind direction but the integral in (\ref{eqn-proccontinuo}) does not have an analytical solution and we resort to the discretized version of the convolution approach to make inference about the process of interest.
In order to compare the proposed models of Sections \ref{sec:Gaussian} and \ref{sec:kernelsdiscretized}, Section \ref{sec:alternative} proposes an alternative parameterization to the one used by \citeasnoun**{PACIOREK2006} when modelling $\Sigma(s)$ in equation (\ref{eq:covarMatern}).

\subsection{Accounting for wind direction in the nonstationary Mat\'ern covariance function}\label{sec:Gaussian}


Here we focus on the nonstationary version of the Matérn covariance function  and introduce the covariate information in the kernel matrices, which we denote as $\Sigma(s,x)$. Let $x(s)=(u(s),v(s))'$ be a vector representing
the directional covariate at location $s\in G$, with $||x(s)||=1$. 
 We propose that the kernel matrices of equation (\ref{eq:covarMatern}) are modelled as
{\small
\begin{equation}
\Sigma(s,x) = \Gamma(s,x)^T \Lambda \Gamma(s,x)=\left[ {\begin{array}{*{20}c}
   {\cos \, \omega(x(s))  } & { - \sin  \, \omega(x(s))  }  \\
   {\sin  \, \omega(x(s))  } & {\, \cos  \, \omega(x(s))  }  \\
\end{array}} \right]\left[ {\begin{array}{*{20}c}
  \lambda_{1}^2   & 0  \\
   0 & \lambda_{2}^2    \\
\end{array}} \right]\left[ {\begin{array}{*{20}c}
   {\cos \, \omega(x(s))  } & {  \sin  \, \omega(x(s))  }  \\
   {- \sin  \, \omega(x(s))  } & {\, \cos  \, \omega(x(s))  }  \\
\end{array}} \right], \label{eq:windgauss}
\end{equation}
}
where $\omega(x(s))  = \arctan \left( {\frac{{v\left( s \right)}}{{u\left( s \right)}}} \right)$. Similar to \citeasnoun**{PACIOREK2006}, each location $s$ has a Gaussian kernel with mean $s$, but we propose the covariance kernel $\Sigma(s,x)$ to vary spatially with the wind direction. The matrix $\Gamma(s,x)$
plays the role of a rotation matrix, whose angle of rotation is given by the arctangent between the components of the wind vector at location $s$. This  makes the Gaussian kernel at location $s$ to coincide with the wind direction at $s$. 
On the other hand, the diagonal matrix $\Lambda$ captures the magnitude of the major and minor axis of the ellipses associated to the contours of the Gaussian kernel. These are assumed fixed across the region $G$. Therefore,  when inference is performed based on $n$ monitoring locations,  the estimated values of $\lambda_{1}^2$  and  $\lambda_{2}^2$ can be interpreted as an overall mean direction of the axis of the ellipses associated with the kernel at each location $s$.
As this process is based on the above matrix kernel we denote it as a {\em Locally Geometric Anisotropic} (LGA) model.

Although this kernel varies with location, it does so according to the variation of $x(.)$. In contrast to  \citeasnoun**{HIGDON1999} and \citeasnoun**{PACIOREK2006}, who make use of independent latent GP, 
we let the directional covariate guide the matrices $\Sigma(s,x)$, reducing significantly the number of parameters to  be estimated.
Clearly, this parameterization of $\Sigma(s,x)$ involves only  three parameters in the specification of $Y(.)$.   

Note that both $\lambda_1^2$ and $\lambda_2^2$ are positive. When performing inference about these quantities we follow the Bayesian paradigm and assign independent, non-informative, inverse gamma prior distributions for these parameters.

\begin{figure}[!htb]
\begin{center}
\subfigure[Wind vectors blowing at the same direction]{\includegraphics[scale=.6]{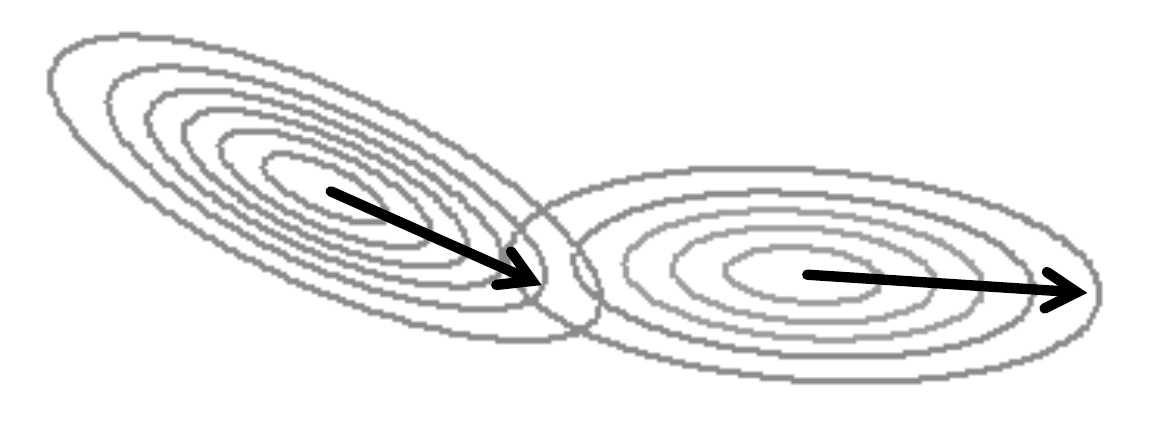}}
\subfigure[Wind vector blowing at opposite directions]{\includegraphics[scale=.6]{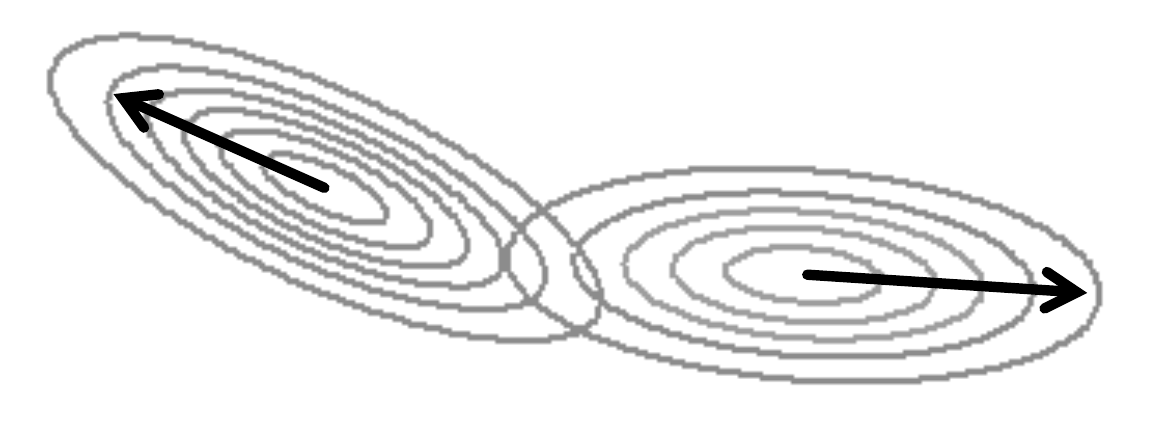}} 
\end{center}
\caption{Illustration of the contours of the kernel matrix proposed in equation (\ref{eq:windgauss}) based on locations with contrasting situations about the wind information (black arrows). \label{fig:ellipses}}
\end{figure}

Although this proposal is efficient in reducing the number of parameters to be estimated and accounting for some wind information, it still fails to compare the directions at which the winds are blowing at two different locations, $s_i$ and $s_j$ in $G$. This undesired feature of this kernel is depicted in the panels of Figure \ref{fig:ellipses}. Note that in both cases the coincident areas of the two ellipses are the same. Therefore, the resultant covariance function, conditional on the same values of $\lambda_1$ and $\lambda_2$, will produce the same values. However, this does not seem reasonable, as the winds in the right panel are blowing in opposite directions. Ideally, we would like to have a correlation function which results in a higher correlation for the wind information in the left panel, when compared to that of the right panel.

In the next section we propose an alternative kernel that makes use of a quantity which captures  how concordant two locations are in terms of the direction at which the wind vectors are pointing at the two locations.

\subsection{An alternative kernel based on a projection measurement}\label{sec:kernelsdiscretized}

We now propose an alternative way  to account for a directional covariate in the kernel function $k_{s,x}(.)$ of equation (\ref{eqn-proccontinuo}). 
We start by proposing a measurement that provides the degree of concordance between any two vectors. This measurement is based on a projection of the mean wind direction between two locations. That is, we say that two vectors, at two distinct locations, tend to be more  concordant, the more similar is the direction in which the respective winds are  blowing.  Next we propose a kernel function that makes use of the norm of this projection measurement. 

\begin{figure}[h]
   \centering
   \includegraphics[width=16 cm,angle=0]{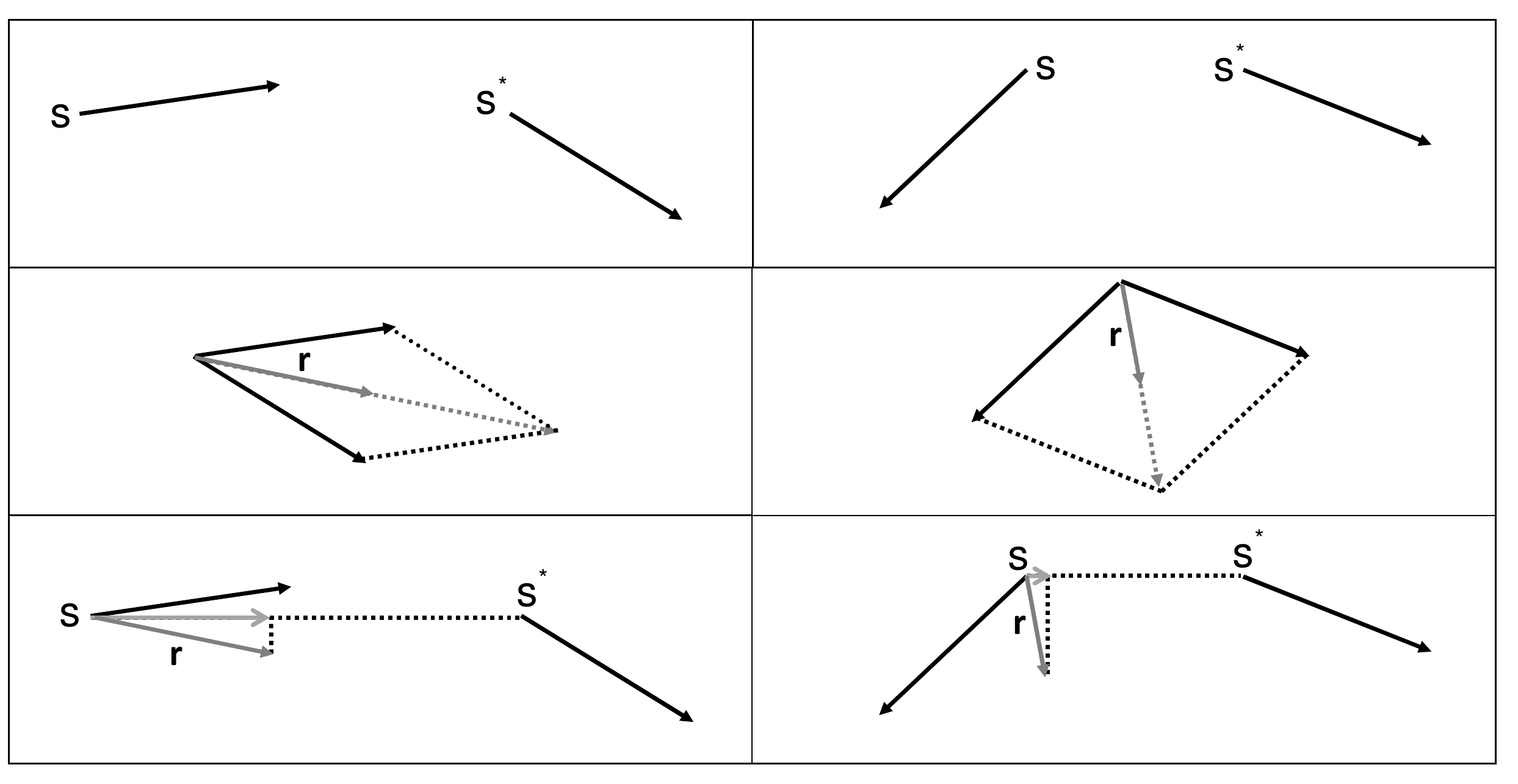}
   \caption{Diagram with directional covariates, black arrows (first row), mean direction $r$ (second row), and resultant projection, gray arrows (third row).}
   \label{fig:projecao}
\end{figure}

\paragraph{ Projection of the average wind direction as a degree of concordance between vectors}
Let $x(s)$ be defined as before, and define $r(x(s),x(s^*))=\frac{(x(s)+x(s^*))}{2}$ as the mean vector when one considers locations $s$ and $s^*$ in $G$. The projection of
the mean vector over the set $\{b\times(s-s^*)+s^*:b \in \reais
\}$ (a straight line that passes through $s$ and $s^*$) is
\begin{equation*}
 proj_x(s,s^*) = {\frac{{\left\langle {r\left( {x(s),x(s^*)}
\right),(s-s^*)} \right\rangle }}{{\left\langle
{(s-s^*),(s-s^*)} \right\rangle }}
\times (s-s^*)},
\end{equation*}
where $\left\langle {a_1,a_2}\right\rangle $ represents the inner product between vectors $a_1$ and $a_2$.
The smaller (greater) the angle between the mean vector and the
direction between the two locations, the greater (smaller) is
$||proj_x(s,s^*)||$, the norm of $proj_x(s,s^*)$. In other words, this measurement takes into account the concordance of the directional covariate and the direction that crosses the locations. Figure \ref{fig:projecao} shows a diagram that depicts how the projection captures the concordance between the vectors $x(s)$ and $x(s^*)$.  In the left column, the measurements  at locations $s$ and $s^*$ are more or less aligned, pointing nearly to the same direction, whereas in the right column, the measurements are nearly perpendicular, pointing to different directions. It is clear from the third row of this picture how the  concordance between the directions is reflected on the norm of the wind mean vector;  that is, $||proj_x(s,s^*)||$ is greater for the vectors on the left panel, than that obtained for the vectors on the right panel of the figure.

Here, the kernel function is modelled as
\begin{equation}\label{eqn-normcontinuo}
k_{s,x}\left( {h} \right) = \frac{{\sigma\alpha_{s,x} \left(h
\right)}}{{\sqrt {\int\limits_G {\alpha_{s,x} \left(h \right)^2 dh}
} }},
\end{equation}
where $\alpha_{s,x}(.)$ is a function that is specified below.
Following the definition of the kernel $k_{s,x}(.)$ in equation (\ref{eqn-normcontinuo}),  $Y(\cdot )$ is  a zero mean Gaussian process with  covariance function
{\small
\begin{eqnarray}\label{eqn-covarqqq}
 \nonumber Cov\left( {Y\left( s \right),Y\left( {s^*} \right)} \right) 
 \nonumber &=& \int\limits_G {k_{s,x}\left( {h} \right)k_{s^*,x^*}\left( h \right)dh} \\
    &=& \sigma^2 \, \, \frac{{\int\limits_G {\alpha_{s,x} \left( {h} \right)\alpha_{s^*,x^*} \left( {h} \right)dh} }}{{\sqrt {\int\limits_G {\left( {\alpha_{s,x} \left( {h} \right)} \right)^2 dh} } \sqrt {\int\limits_G {\left( {\alpha_{s^*,x^*} \left( {h} \right)} \right)^2 dh} }
  }} \, \, .
\end{eqnarray}
}
Now the reason for the form of the kernel proposed in equation (\ref{eqn-normcontinuo}) is clear. The covariance function between any locations $s$ and $s^*$ can be expressed as a product of the variance $\sigma^2$, by a correlation function.  This decomposition of the covariance function is a standard approach in the geostatistical literature, see e.g. \citeasnoun**{BANERJEE2004}. Therefore,  
$\sigma$ can be viewed as the standard deviation of $Y(.)$, and the function
$\alpha_{s,x}(.)$ is related to the spatial correlation of the process.

As previously mentioned, the kernel function is any nonnegative function which usually decays monotonically from its mode. The motivation for the way we model $\alpha_{s,x}(.)$ comes from the exponential correlation function. We assume
\begin{equation}\label{eqn-ventoedist}
  \alpha_{s,x}(h) = \left\{ {\begin{array}{*{20}c}
   {\exp \left( { - \frac{{\left\| {s - h } \right\|}}{{\phi _1  + \phi _2 \left\| {proj_x\left( {h} \right)} \right\|}}} \right),\text{ if }s \ne h } \hfill  \\
   {1,\text{ if }s = h }. \hfill  \\
\end{array}} \right.
\end{equation}
In order to guarantee an exponential decay of the kernel we assume $\phi_{1}>0$ and $\phi_{2}>0$.
This kernel decays exponentially with the ratio between the Euclidean distance between $s$ and $h$, and the norm of the projection of the average wind direction. The influence of the different components is captured by two parameters, $\phi_1$ and $\phi_2$. In particular,  $\phi_{1}$ measures the effect of the Euclidean distance, and $\phi_{2}$ the effect of the directional covariate in the correlation structure of $Y(.)$. From equation  (\ref{eqn-ventoedist}) we see that isotropy is approximately attained  when $\phi_{2} \rightarrow \, 0$. We denote a process based on this  kernel $\alpha_{s,x}(.)$ as  a \textit{Projection Model}. 

In Section \ref{sec:dataana}, when fitting this model to the ozone data, we assign independent, non-informative, inverse gamma prior distributions for $\phi_1$ and $\phi_2$.

The kernel in (\ref{eqn-ventoedist})  does not provide a closed form of the covariance structure of $Y(.)$. For this reason, we resort to the discretized version of this process and assume $Y(s)=\sum_{l=1}^m k_{s,x}(h_l)W(h_l)$. Therefore, all the integrals in equation (\ref{eqn-covarqqq}) are substituted by their discretized versions. Note that similarly to the kernel parameterization proposed in Section \ref{sec:Gaussian}, the number of parameters to be estimated associated to $Y(.)$ under  equation (\ref{eqn-ventoedist}), are significantly smaller when compared to the general approaches reviewed in Section \ref{sec:convol}. Because of the reduced number of parameters to be estimated, when fitting this model in the analysis presented in Section \ref{sec:dataana},  we consider the monitoring locations as the grid points to approximate the integrals of equation (\ref{eqn-covarqqq}). We discuss the selection of the grid points in further detail in Section \ref{sec:conclusao}. Section A of the Supplementary Materials depicts the behaviour of the proposed correlations functions for different values of the parameters.

\subsection{An alternative model for $\Sigma(s)$ of the nonstationary Mat\'ern covariance function \label{sec:alternative}}

Sections \ref{sec:Gaussian} and \ref{sec:kernelsdiscretized} discuss different ways of accounting for wind direction in the covariance structure of $Y(.)$ resulting in nonstationary covariance functions. It is reasonable to compare this approach with those proposed by \citeasnoun**{HIGDON1999} or \citeasnoun**{PACIOREK2006}, which are able to capture  more flexible covariance structures. 
In Section \ref{sec:dataana} we tried fitting both models to the ozone data of Section \ref{sec:motivation}, but did not get convergence of the chains. Both models heavily depend on pre-fixed constants; and, from our experience, the algorithms seem to be quite sensitive to the choice of these constants.

Following \citeasnoun**{PACIOREK2006}, we construct each $\Sigma(s)$ in equation (\ref{eq:covarMatern}) using the spectral decomposition, that is, we assume  $\Sigma \left( s \right) = \Gamma \left( s \right)^T\Lambda \left( s \right)\Gamma \left( s \right)$. The matrix $\Gamma(s)$ is a diagonal matrix of eigenvalues,  $\Lambda \left( s \right) = \left[ {\begin{array}{*{20}c}
   {\lambda _1 \left( s \right)} & 0  \\
   0 & {\lambda _2 \left( s \right)}  \\
\end{array}} \right]$. Different from  \citeasnoun**{PACIOREK2006}, we assume $\Gamma \left( s \right) = \left[ {\begin{array}{*{20}c}
   {\cos \left( {\theta \left( s \right)} \right)} & {\sin \left( {\theta \left( s \right)} \right)}  \\
   { - \sin \left( {\theta \left( s \right)} \right)} & {\cos \left( {\theta \left( s \right)} \right)}  \\
\end{array}} \right]$. 

The kernel matrices are expected to vary smoothly across the spatial domain.  In our proposal this is guaranteed through the prior specification of $\lambda_j(s)$, for $j=1,2$, and $\theta(s)$. We assume $\log \lambda_j(.)$ follow independent Gaussian processes, with mean $\mu_{\lambda_j}$, variance $\sigma_\lambda^2 $ and a squared exponential covariance function, that is, $Cov(\log \lambda_j(s),\log \lambda_j(s^*))=\sigma_{\lambda}^2 \, \exp\left\{-\left(\frac{||s-s^*||}{\phi_\lambda}\right)^2\right\}$.
Next, we model $\theta(s)$ as 
\begin{equation}
  \theta(s)=\frac{\pi}{2} \, \Phi(\gamma(s)), \label{eq:modeltheta}
\end{equation}
where $\Phi(.)$ denotes the cumulative distribution function of the standard normal distribution. We assume $\gamma(s)$ follows a Gaussian process with mean $\mu_{\gamma}$ and a squared exponential covariance function, that is, $Cov(\gamma(s),\gamma(s^*))=\sigma_\gamma^2 \exp\left\{-\left(\frac{||s-s^*||}{\phi_\gamma}\right)^2\right\}$. 

Note that $\theta(s)$ varies smoothly across the region induced by the GP for $\gamma(s)$. Under the parametrization in equation (\ref{eq:modeltheta}), $\theta(s) \in [0,\pi/2]$. This restriction on the values of the rotation angles is imposed to avoid possible unidentifiability problems due to the specification of $\Gamma(s)$.

When fitting this model in Section \ref{sec:dataana},
we assume the following prior specification: independent zero mean normal prior distributions, with large variance, for $\mu_{\lambda_1}$ and $\mu_{\lambda_2}$, and a standard normal prior distribution for $\mu_{\gamma}$. For the decay parameters $\phi_{\lambda}$ and $\phi_{\gamma}$ we assign an inverse gamma prior distribution with parameters fixed based on the idea that the practical range (when the correlation is equal to 0.05) is reached at half of the maximum distance between geographical locations and the prior variance is fixed at some relatively large value. For the variance parameters, $\sigma^2_{\lambda}$ and $\sigma^2_\gamma$, we assign an inverse gamma prior distribution with mean and variance equal to $1$.

\section{Inference procedure \label{sec:inference}}

Assuming the model proposed in equation (\ref{eqn:geral}), and letting ${\bf z}=(z(s_1),\cdots,z(s_n))'$ be a partial realization from a Gaussian process with mean vector ${\bs \mu}$ and covariance matrix $\Sigma$, the likelihood function is given by
\begin{equation*}
L({\bf z};{\bs \theta})=(2\pi)^{-n/2} \mid \Sigma \mid ^{-1/2}
\exp\left\{-\frac{1}{2} ({\bf z}-{\bs \mu})^T \Sigma^{-1}({\bf z}-{\bs \mu})
\right\}.
\end{equation*}
Following \citeasnoun**{SANSOGUENNI2004}, we write the likelihood function in terms of $\eta=\tau^2 / \sigma^2$, such that $\Sigma=\tau^2(I_n+\eta^{-1} \, \Omega(\bs \delta))$, where $I_n$ represents a $n$-dimensional identity matrix, ${\bs \delta}=(\delta_1,\delta_2,\cdots,\delta_k)$ comprises the parameter vector of the kernel function, and $\Omega({\bs \delta})$ is the associated resultant covariance matrix. Then the parameter vector to be estimated is given by $\bs \theta=(\bs \mu,\tau^2,\eta,\bs \delta)$.

Inference is performed under the Bayesian paradigm, and we assume the components of the parameter vector ${\bs \theta}$ to be independent {\em a priori}. Generally, one assumes $\mu(s)={\bs Q}(s)' {\bs \beta}$, where ${\bs Q}(\cdot )$ is a $p$-dimensional vector with covariates that might influence the mean of $Z(\cdot )$, and ${\bs \beta}$ is a $p$-dimensional parameter vector. It is common practice to assign independent zero mean normal prior distributions with some fixed large variance to each  $\beta_j$, $j=1,\cdots,p$. For $\tau^2$ and $\eta$ we assign independent, non-informative, inverse gamma prior distributions with parameters equal to 0.1. The elements of ${\bs \delta}$ are model dependent. The Supplementary Material discusses the prior specification of the parameters in $\bs \delta$ under each of the models proposed in the previous section.

Regardless of the fitted model, the resultant posterior distribution, $\pi({\bs \theta} \mid {\bf z})$, does not have a closed form, and we resort to MCMC algorithms to obtain samples from the target distribution. In particular, we make use of the Gibbs sampler with some steps of the Metropolis-Hastings algorithm. The full conditional posterior distributions for each of the models are shown in Section 2 of the Supplmentary Material.

\subsection{Predictive inference \label{sec:predictive}}

We now discuss how to obtain predicted values at unmonitored locations of interest. Let ${\bf{z}}^* = \left(z(s_{n+1}), ..., z(s_{n+q})\right)'$ be a vector representing the value of the process at $q$ unmonitored locations of interest. Samples $Z(s)$ are being generated from the multivariate normal distribution,
$N({\bs \mu},{\Sigma})$, the posterior predictive distribution,
$p({\bf z}^{*}|{\bf z})$, is given by
\begin{eqnarray}
p({\bf z}^{*}|{\bf z})=\int_{\bs \theta} p({\bf
z}^{*}|{\bf z},{\bs \theta})\pi({\bs \theta}|{\bf z})
d{\bs \theta}. \label{eq:preditiva}
\end{eqnarray}
From the theory on the multivariate normal distribution,
the joint distribution of $\left({\bf{Z}},{\bf{Z}}^*\right)$, conditional on $\bs \theta$, follows a multivariate normal distribution,
\begin{equation*}
\left( {\left. { {\begin{array}{*{20}c}
   \bs Z \\
   {\bs Z^*}  \\
\end{array}}} \right|{\bs \theta} } \right)\sim N\left( {\left[ {\begin{array}{*{20}c}
   {{\bs \mu}_A }  \\
   {{\bs \mu}_B }  \\
\end{array}} \right],\left[ {\begin{array}{*{20}c}
   {\Sigma _{AA} } & {\Sigma _{A,B} }  \\
   {\Sigma _{BA} } & {\Sigma _{BB} }  \\
\end{array}} \right]} \right).
\end{equation*}
Therefore, based on the properties of the partition of the multivariate normal distribution,
\begin{equation}\label{eqn:predict}
\left( {{\bs Z}^* |{\bs z},{\bs \theta} } \right)\sim N\left( {{\bs \mu} _B  + \Sigma _{BA} \Sigma _{AA} ^{ - 1} \left( {{\bs z} - {\bs \mu}_A } \right),\Sigma _{BB}  - \Sigma _{BA} \Sigma _{AA} ^{ - 1} \Sigma _{AB} } \right).
\end{equation}

The integration in (\ref{eq:preditiva}) does not have an analytical
solution, however approximations can be easily obtained through
Monte Carlo methods \cite**{GAMERMAN2006}. For each sample
$l$, $l=1,\cdots, L$, obtained from the MCMC algorithm, we can
obtain an approximation for (\ref{eq:preditiva}), by sampling from the
distribution in (\ref{eqn:predict}) and computing
\begin{eqnarray}
p({\bf z}^{*} | {\bf z}) \approx \frac{1}{L}\sum_{l=1}^L \,
p({\bf z}^{*} | {\bs \theta}^l). \label{eq:predictivelik}
\end{eqnarray}
The approximation above is also suitable for model comparison.
Usually, one holds  a set of observations out from the inference procedure, say ${\bf z}^{*}$,
and uses the predictive likelihood to compare all fitted models to check under which model the actual observed values are more likely to be generated from. Greater values of $p({\bf z}^{*} | {\bf z})$ in (\ref{eq:predictivelik}) point to the best model.


\paragraph{Interpolating the wind field} 

Both models proposed in Sections \ref{sec:Gaussian} and \ref{sec:kernelsdiscretized}, require the measurements of the  components of wind direction, $x(.)$ at the same locations where the process $Z(.)$ is observed. The approach proposed in Section \ref{sec:Gaussian} requires that $x(.)$ and $Z(.)$ are observed at the same locations. On the other hand,  the parameterization of the kernel in equation (\ref{eqn-ventoedist}) requires that the components of $x(.)$ are observed at the same locations of the grid points, $\{h_1,\cdots, h_m\}$.  If these observations are not available we propose to model $x(.)$ and obtain interpolated values for unmonitored locations of interest.

We suggest that spatial interpolation of $x(s)$ follows  \citeasnoun**{WIKLE2001} who, based on physical grounds, assume prior independence between the components $u(s)$ and $v(s)$ of $x(.s)$.
Therefore, we assume that the first coordinate of the vector $x(.)$, follows a Gaussian process $\{U(s):s\in G\}$, with mean $\mu_{u}$ and exponential covariance function, $Cov(U(s),U(s^*))=\sigma_u^2 \exp\left(-\frac{||s-s^*||}{\phi_u}\right)$ $\forall s,s^*\in G$. And  the second coordinate of the vector $x(s)$ follows another, conditionally independent, Gaussian process $\{V(s):s\in G\}$, with mean $\mu_{v}$ and exponential covariance function, $Cov(V(s),V(s^*))=\sigma_v^2 \exp\left(-\frac{||s-s^*||}{\phi_v}\right)$ $\forall s,s^*\in G$. We fit these independent models to each of the components of the wind field, and obtain estimated  values of $u(\cdot )$ and $v(\cdot )$ at all  unmonitored locations of interest. The inference procedure for the parameters in the model for $Z(.)$ follow conditioned on these interpolated values.


\subsection{Model Comparison \label{interpolation}}

In Section \ref{sec:dataana} we fit different models to the ozone levels described in Section \ref{sec:motivation}. We briefly review two model comparison criteria: the Deviance Information Criteria (DIC), the Posterior
Predictive  Loss Criterion. See \citeasnoun**{BANERJEE2004} for more discussion about these criteria.

\paragraph{Deviance Information Criterion}

\citeasnoun**{DIC2002} propose a
generalization of the AIC based on the posterior distribution of
the deviance, $D({\bs \theta})=-2\log L({\bf z};{\bs \theta})$. The
Deviance Information Criterion (DIC) is defined as
$$
DIC=\overline{D}+p_D=2\overline{D}-D(\overline{\bs \theta}),
$$
where $\overline{D}$ defines the posterior expectation of the
deviance, $\overline{D}=E_{{\bs \theta} \mid {\bf z}}(D)$, and $p_D$
is the effective number of parameters,
$p_D=\overline{D}-D(\overline{\bs \theta})$ and here
$\overline{\bs \theta}$ represents the posterior mean of the
parameters. Smaller values of DIC indicate a better fitting model.
Note that computation of DIC is easily achieved through MCMC
methods. 

\paragraph{Posterior Predictive Loss Criterion}

An alternative to DIC is the posterior predictive loss (PPL) introduced
by \citeasnoun{EPD1998}. This measurement is based on
replicates of the observed data, $Z_{l,rep} \, l=1,\cdots, n$, and
the selected models are those that perform well under a so-called
loss function. This loss function penalizes actions both for
departure from the corresponding observed value as well as for
departure from what we expect the replicate to be
\cite{BANERJEE2004}. The criterion is computed via
\begin{eqnarray*}
&&D_k=\frac{k}{k+1}G+P, 
\end{eqnarray*}
where $G=\sum_{l=1}^n (\mu_l-z_{l,obs})^2$ and $P=\sum_{l=1}^n \sigma_l^2$.
Here $\mu_l=E(Z_{l,rep} \mid {\bf z})$ and
$\sigma_l^2=Var(Z_{l,rep} \mid {\bf z})$, denote the mean and variance of
the predictive distribution of $Z_{l,rep}$ given the observed data
${\bf z}$. \citeasnoun{EPD1998} mention that
ordering of models is typically insensitive to the choice of $k$,
therefore we fix $k=1$. Notice that at each iteration of the MCMC
we can obtain replicates of the observations given the sampled
values of the parameters.


\section{Data Analysis \label{sec:dataana}}

In this section, we illustrate the proposed models using measurements of ozone observed at 3pm, on November, 11th, 2008. The monitoring locations and the observed values are shown in Figure \ref{fig:data}. We fit  5 different models to this dataset. The aim is to compare how the different models behave in terms of model fitting, spatial interpolation and, specially, the associated uncertainty of the spatial interpolation.


\paragraph{Fitted Models}
 The notation of the fitted models is the following:
\begin{itemize}
\item[M1] Isotropic model with Mat\'ern covariance function
\item[M2] Elliptical anisotropic model with Mat\'ern covariance function, that is
\begin{equation*}
  Cov(Y(s_i),Y(s_j))= \sigma^2 \left( {2^{\nu  - 1} \Gamma \left( \nu  \right)} \right)^{ - 1} \left( {\frac{\sqrt{\left\| u \right\|^T \Lambda  \left\| u \right\|}}{\varphi }} \right)^\nu \kappa_\nu  \left( {\frac{\sqrt{\left\| u \right\|^T \Lambda  \left\| u \right\|}}{\varphi }} \right),
\end{equation*}
where
\begin{equation*}
\Lambda  = \left[ {\begin{array}{*{20}c}
   {\cos \, \theta  } & { - \sin \, \theta }  \\
   {\sin \, \theta  } & {\cos \, \theta  }  \\
\end{array}} \right]\left[ {\begin{array}{*{20}c}
   {\lambda _1 } & 0  \\
   0 & {\lambda _2 }  \\
\end{array}} \right]\left[ {\begin{array}{*{20}c}
   {\cos \, \theta  } & {\sin \, \theta }  \\
   { - \sin \, \theta  } & {\cos \, \theta  }  \\
\end{array}} \right],
\end{equation*}
where $0<\theta<\pi/2$, because of unidentifiability reasons, $\kappa_\nu(.)$ denotes the modified Bessel function of the third type and order $\nu$, $\Gamma$ is the usual Gamma function, $\varphi>0$ is a parameter related to the decay of the correlation as the distance between the locations increases and $\nu>0$ determines the smoothness of the process. M1 is a particular case of M2 when  $\theta=0$ and  $\lambda_1=\lambda_2=1$.
\item[M3] Nonstationary Mat\'ern covariance function with $\Sigma(s,x)$ as in equation (\ref{eq:windgauss})
\item[M4] Covariance function based on the Projection model as presented in  equation (\ref{eqn-covarqqq})
\item[M5] Nonstationary Mat\'ern covariance function with $\Sigma(s)$ as proposed in Section \ref{sec:alternative}
\end{itemize}
For all models we fixed the smoothness parameter of the Mat\'ern covariance function ($\nu$) at $1$. For models M1 to M4 we let the MCMC run for 50,000 iterations, considered 10,000 as burn in,  and kept every 10-th iteration, to reduce autocorrelation within the chains. For model M5 we let the chain run for 700,000 iterations, considered 100,000 as burn in, and kept every 100-th iteration. 
Table \ref{tab:time} shows the computational time needed to run each of the models in an Intel(R) Core(TM)2 Quad CPU Q9550 2.83GHz computer with 4 GB of RAM.  Although M5 is a more flexible model, as it is able to capture any source of nonstationarity in the process, it requires a much longer chain to attain convergence. Moreover, when implementing the MCMC for model M5, it is really challenging to tune the variance of the proposal distributions of the parameters involved in the modelling of $\Sigma(s)$ as proposed in Section \ref{sec:alternative}.
\begin{table}[!hb]
\begin{centering}
\begin{tabular}{|ccc|}
\hline 
 & Computational & No. Iterations \tabularnewline
Model &  time & minute \tabularnewline
\hline 
M1 & 7 min & 7142.85\tabularnewline
M2 & 1 h 26 min & 581.39\tabularnewline
M3 & 1 h 55 min & 434.78\tabularnewline
M4 & 21 min & 2380.95\tabularnewline
M5 & 144 h 30 min & 80.73\tabularnewline \hline
\end{tabular}
\par\end{centering}
\caption{Computational time, and number of iterations per minute, to run the MCMC algorithm for 50,000 iterations for  models M1, M2, M3, M4, and 700,000 iterations for model M5 in an Intel(R) Core(TM)2 Quad CPU Q9550 2.83GHz computer with 4 GB of RAM.} \label{tab:time}
\end{table}

Table \ref{tab:comparison} shows the values of four different model comparison criteria, DIC, PPL, the predictive likelihood based on the circled locations shown in Figure \ref{fig:data}, and the mean squared error (MSE). The MSE was computed using the mean of the posterior predictive distribution as the fitted value.
Clearly, models M1 and M2 result in the worst performance in terms of PPL and DIC. This is an indication that the assumption of isotropy or elliptical anisotropy is not reasonable for this dataset.  On the other hand, M4 results in the smallest values of PPL and DIC, followed by model M5, and  M3. When considering the predictive likelihood based on the locations left out from the inference procedure, model M5 does not perform as well as models M3 and M4. But this result might be sensitive to the set of locations which were left out from the inference procedure.
\begin{table}[!h]
\begin{tabular}{|c|ccc|ccc|c|c|}
\hline 
\multirow{2}{*}{Model} & \multicolumn{3}{c|}{PPL} & \multicolumn{3}{c|}{DIC} & \multirow{1}{*}{Predictive} & \multirow{2}{*}{MSE}\tabularnewline
\cline{2-7} 
 & G & P & $D_1$ & $\overline{D}$ & $p_{D}$ & DIC & likelihood & \tabularnewline \hline
M1 & 458.77 & 1395.79 & 1625.18 & 326.06 & 3.04 & 329.10 & $2.33 \times 10^{-08}$ & 131.10\tabularnewline
M2 & 251.37 & 1050.26 & 1175.95 & 318.70 & -1.80 & 316.37 & $6.18 \times 10^{-08}$  & 100.53\tabularnewline
M3 & 87.80 & 639.06 & 682.96 & 309.46 & 3.88 & 313.34 & $6.49 \times 10^{-07}$ & 45.33\tabularnewline
M4 & 90.38 & 464.93 & 510.12 & 289.79 & 4.03 & 293.82 & $4.93 \times 10^{-06}$ & 25.80\tabularnewline
M5 & 59.97 & 539.60 & 569.59 & 302.29 & 3.03 & 305.32 & $9.98 \times 10^{-08}$ & 70.46\tabularnewline \hline
\end{tabular}
\caption{Model comparison criteria: PPL, DIC, the predictive likelihood based on the circled locations in Figure \ref{fig:data}, and the mean squared error, under each fitted model.\label{tab:comparison}}
\end{table}

\begin{figure}[!hbt]
\begin{center}
\subfigure{\includegraphics[scale=.45]{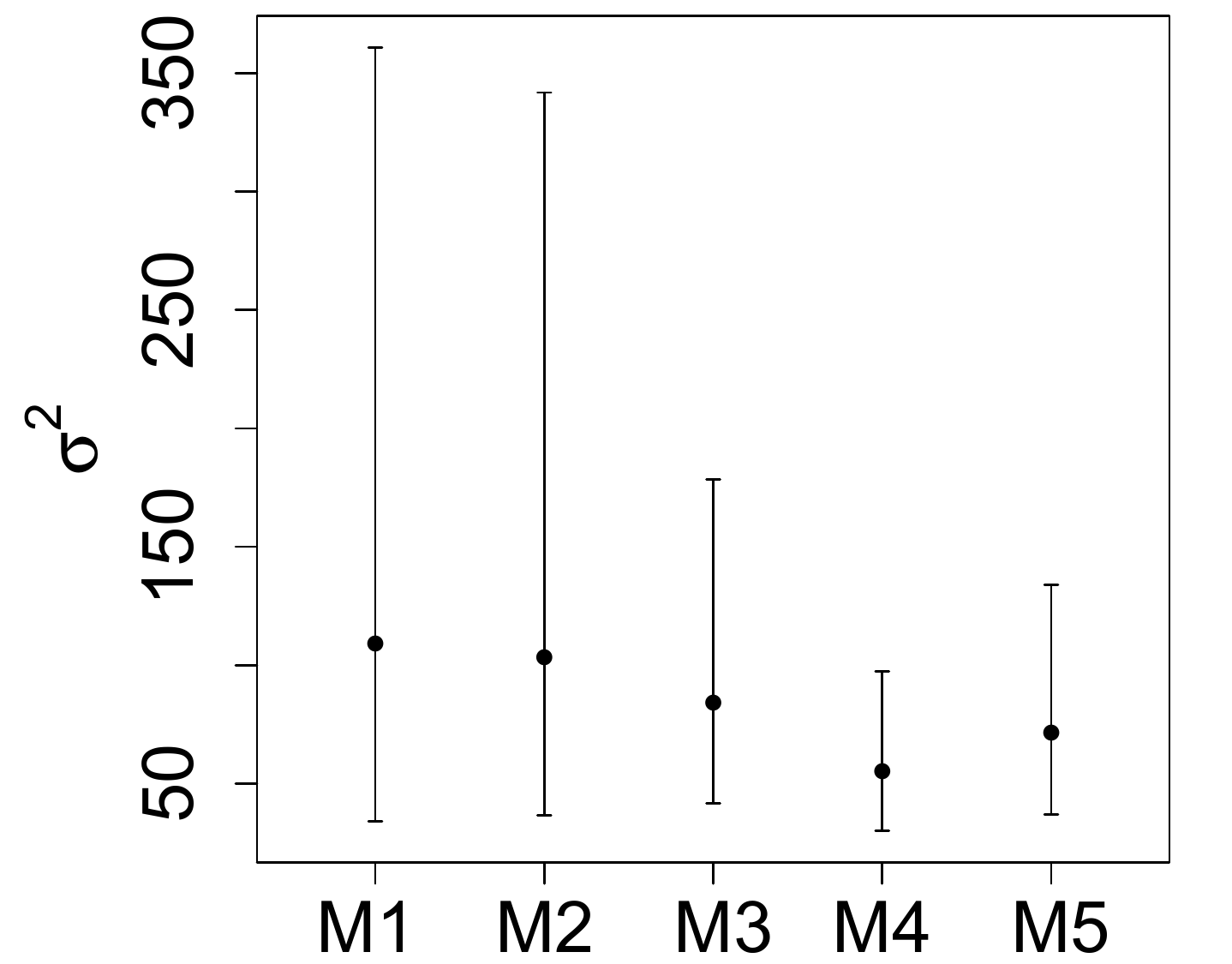}}
\subfigure{\includegraphics[scale=.45]{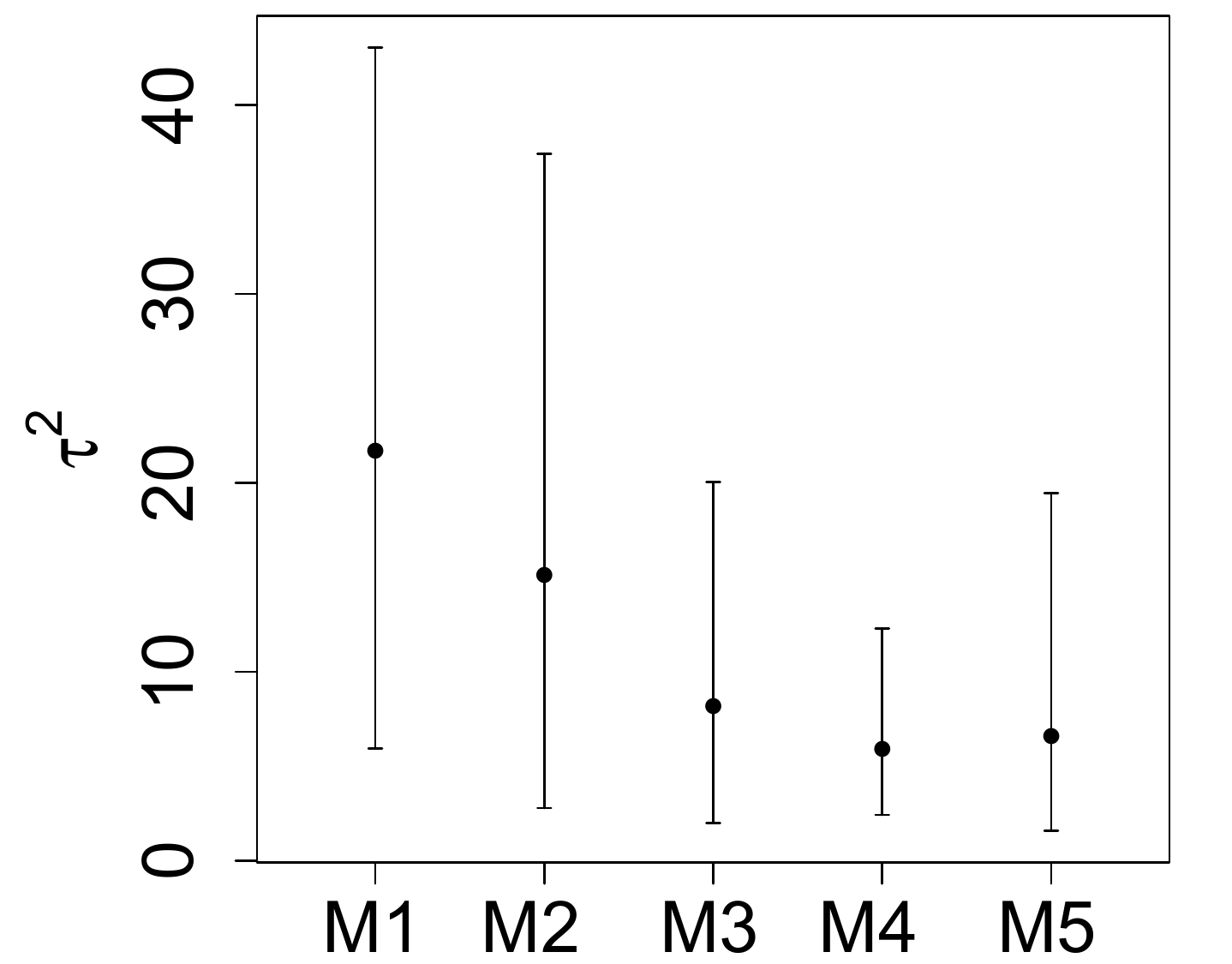}}
\end{center}
\caption{Posterior summary of the variance of the spatial process ($\sigma^2$) and the nugget effect ($\tau^2$) under each of the fitted models M1-M5. \label{fig:variance}}
\end{figure}

Figure \ref{fig:variance} presents the posterior summary of the variance components under each fitted model. The aim is to compare the estimated values of the variance of the spatial component $\sigma^2$ and the nugget effect, $\tau^2$. Clearly, the models with more flexible covariance functions, M3, M4 and M5, result in the smallest values of the nugget effect. And in particular, model M4 results in the smallest value of $\tau^2$. This suggests that there is less structure left in the measurement error under model M4 when compared to the other models. On the other hand, models M1 and M2, the simplest ones, result in the 
largest values of $\tau^2$, suggesting that there is some structure left in the data that the spatial component of these respective models is not able to capture.

\begin{figure}[!h]
\begin{center}
\subfigure[Estimated ellipses under M3]{\includegraphics[scale=.5]{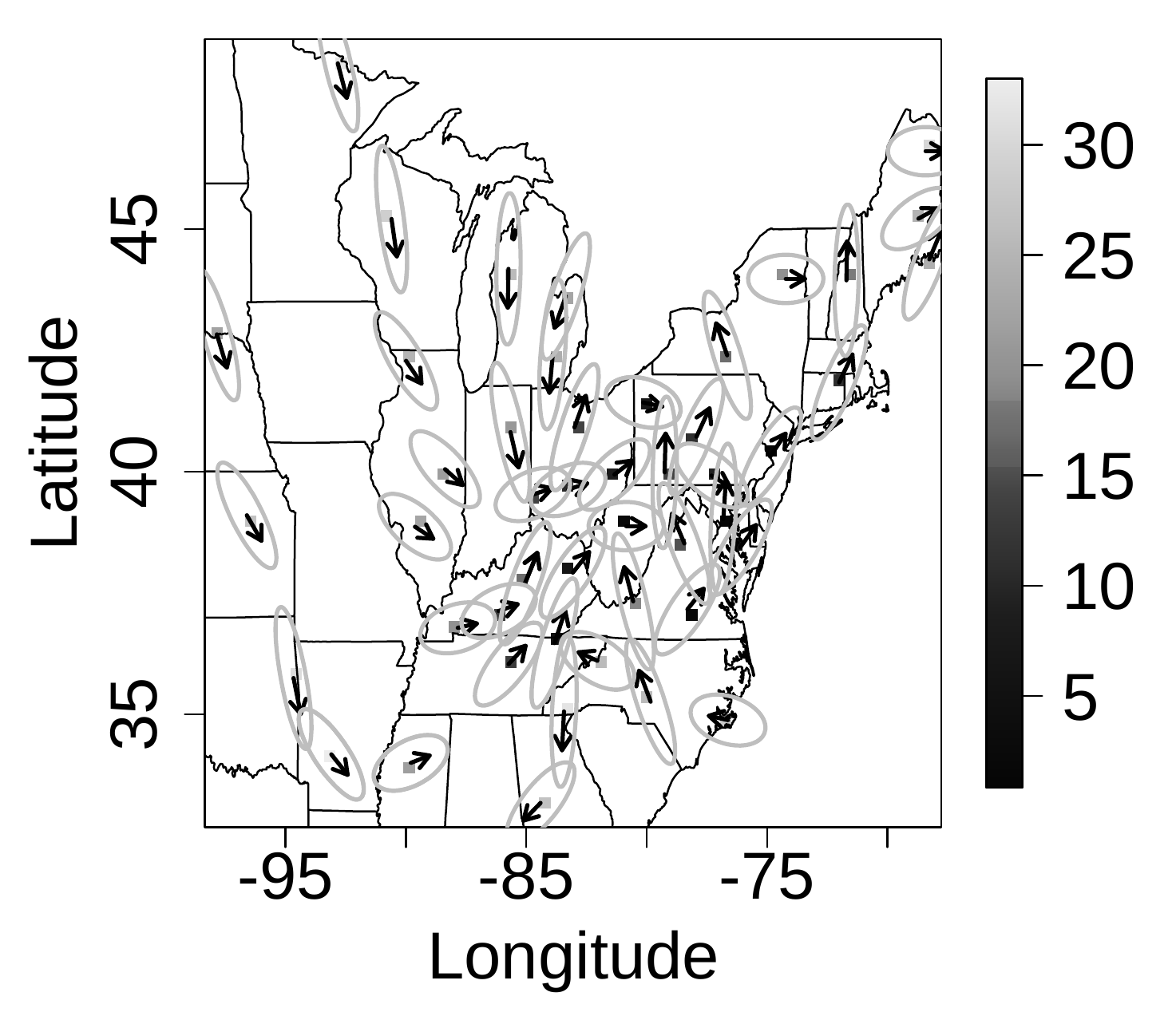}}
\subfigure[Estimated ellipses under M5]{\includegraphics[scale=.5]{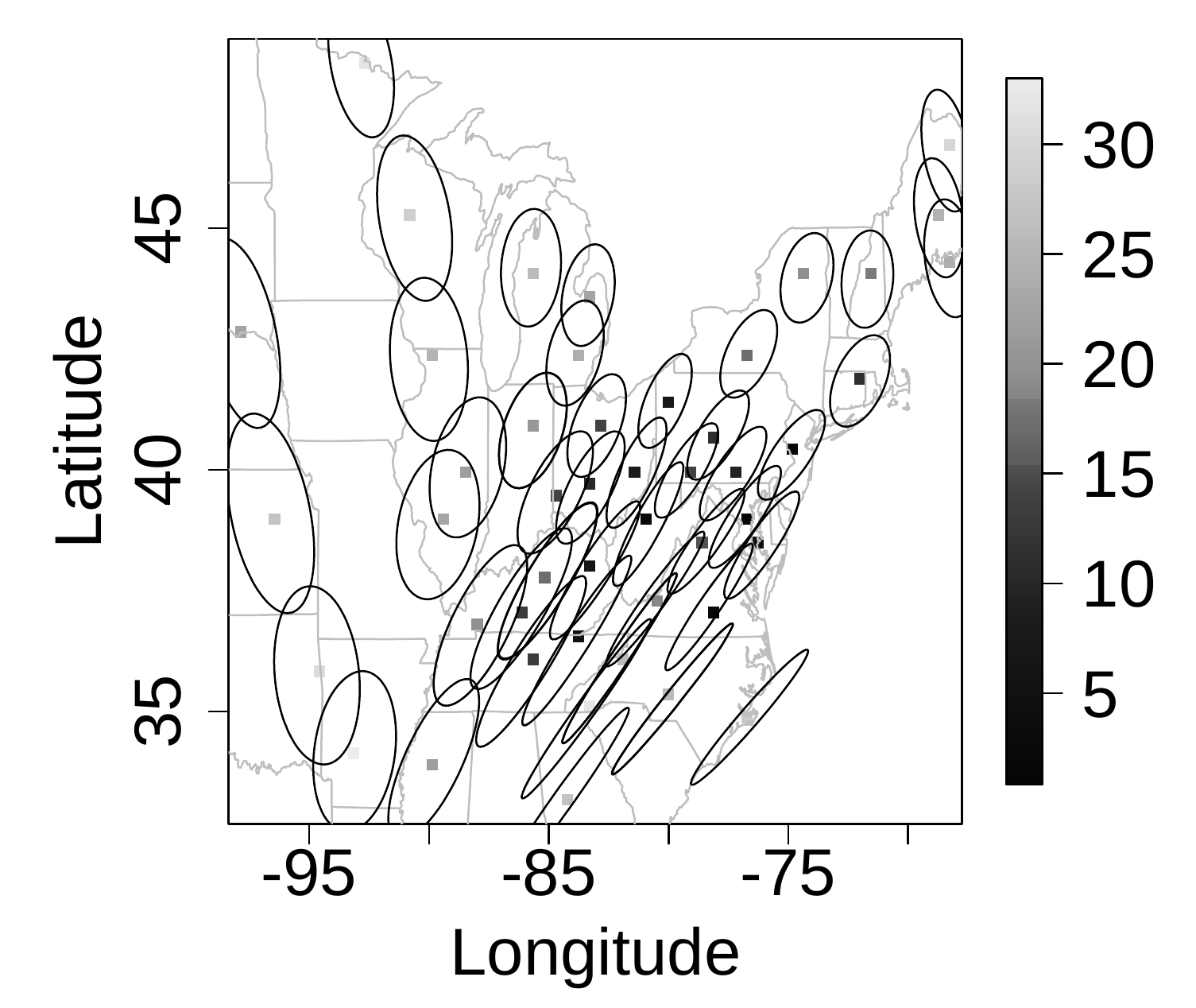}}
\caption{Left panel: wind vectors and respective estimated ellipses under model M3.  Right panel: Estimated ellipses under model M5. In both panels the gray scale is associated with the observed values of ozone. The ellipses were scaled to ease visualization.\label{fig:ellipse}}
\end{center}
\end{figure}

Figure \ref{fig:ellipse} shows the posterior mean of the resultant ellipses obtained under models M3 and M5. This helps to understand how the spatial correlations are changing across the region under  models M3 and M5. Clearly, the ellipses under model M3 follow the direction of the wind field.
Although model M5 does not use any information about the wind field, at some parts of the region, the  direction of the ellipses resemble the observed wind field (e.g. see the west and the middle portions of the region). 

\begin{figure}[!hbt]
\begin{center}
\subfigure{\includegraphics[scale=.4]{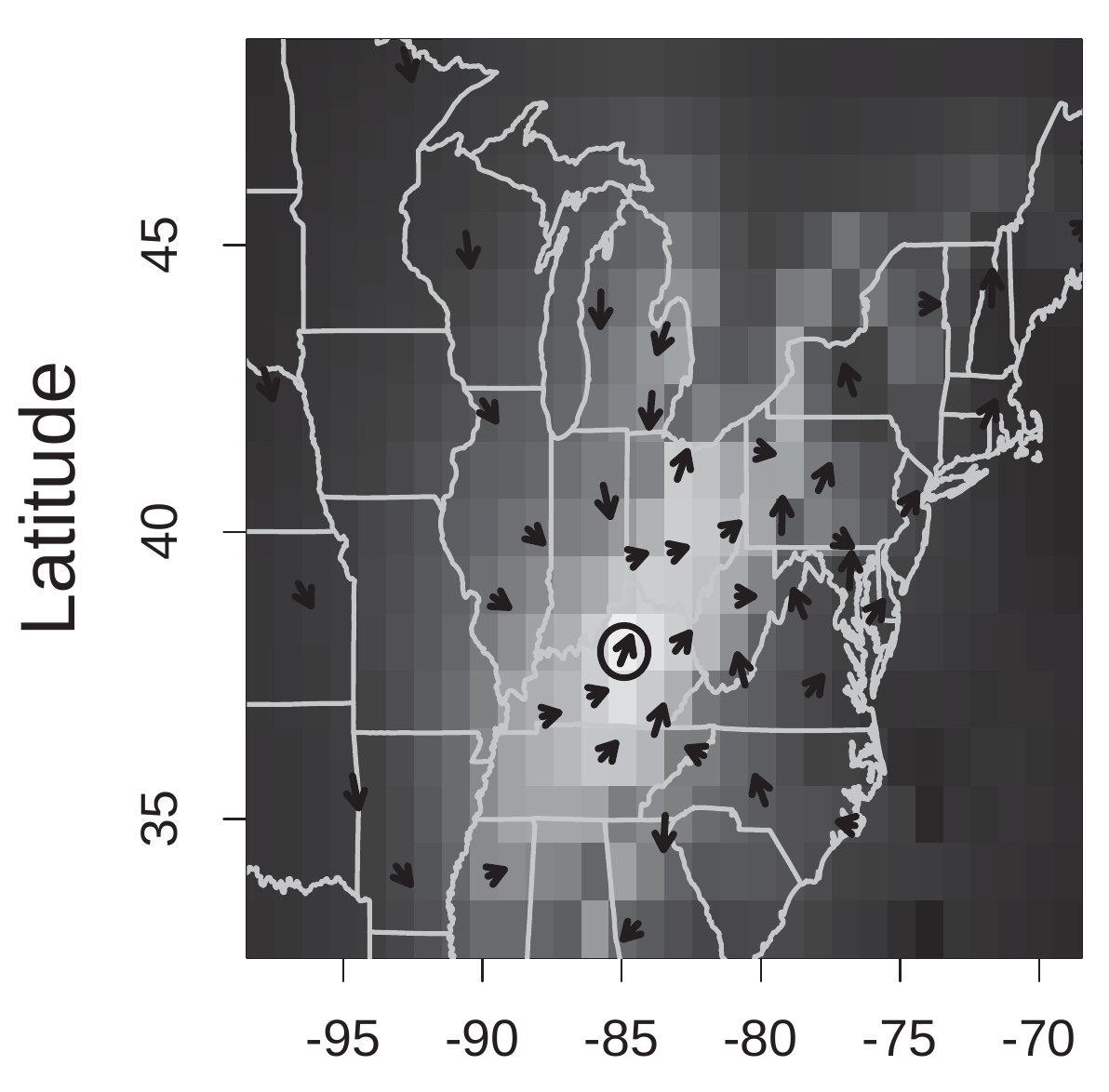}}
\subfigure{\includegraphics[scale=.4]{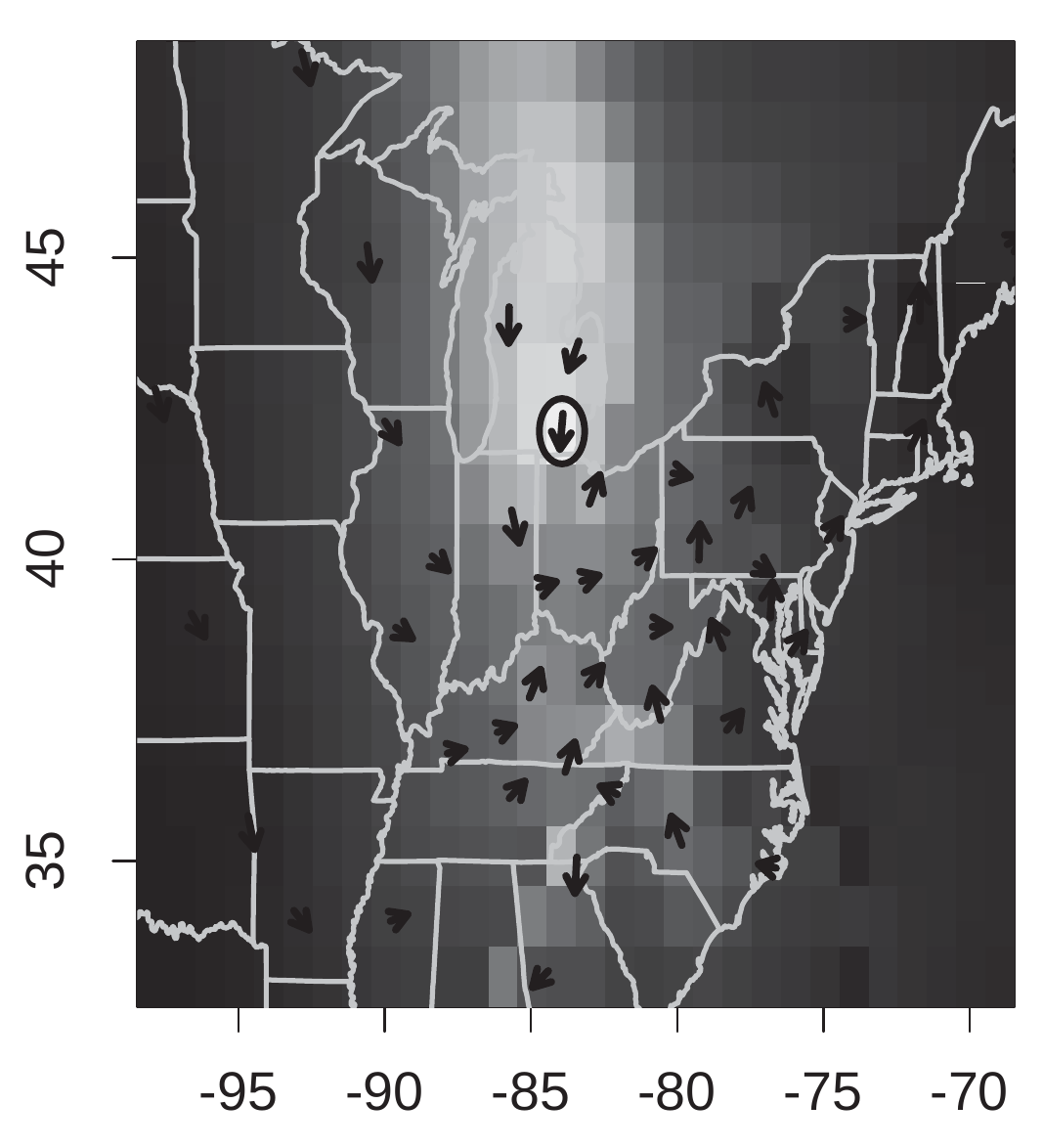}}
\subfigure{\includegraphics[scale=.4]{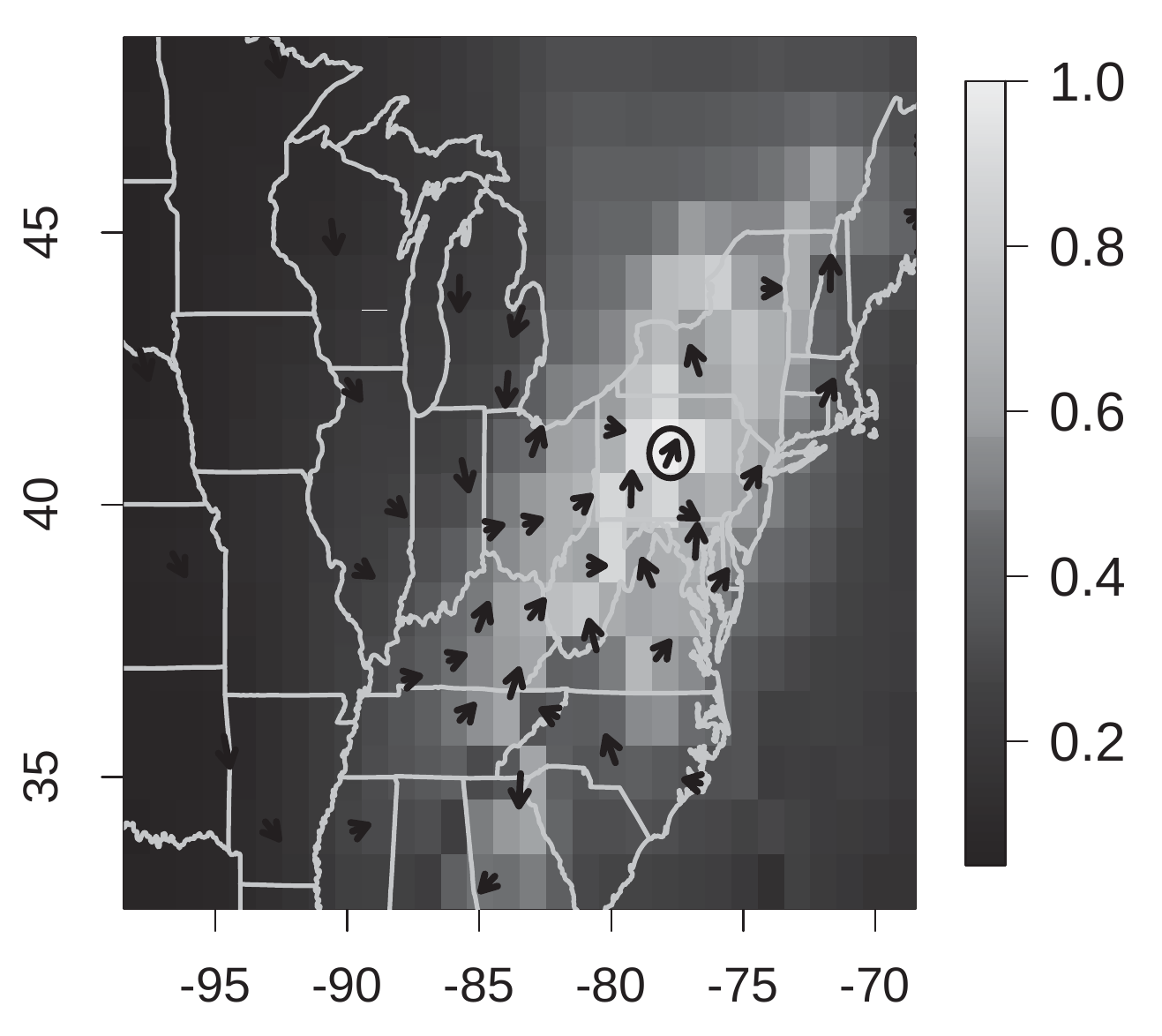}}
\setcounter{subfigure}{0}
\subfigure{\includegraphics[scale=.4]{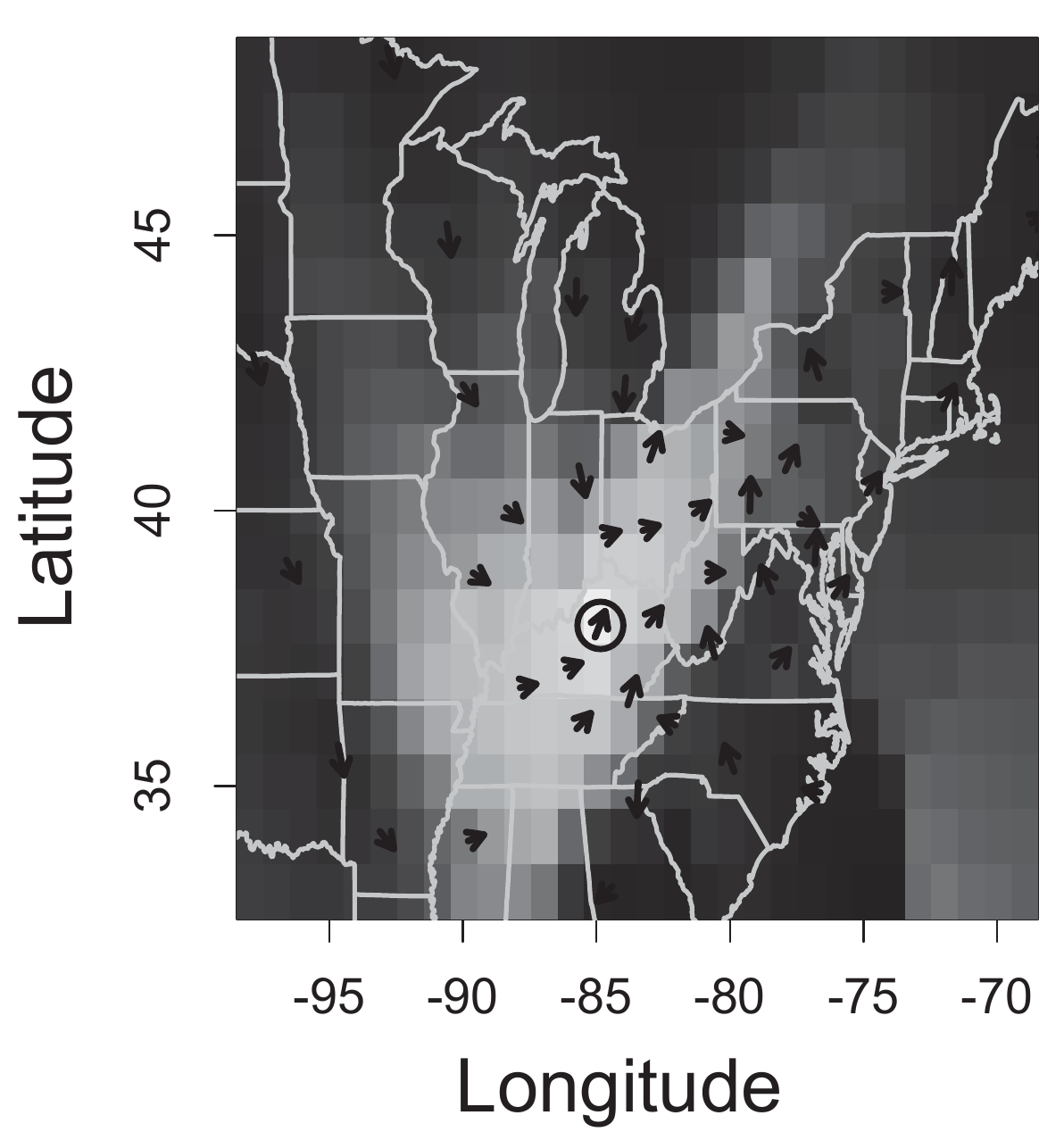}}
\subfigure{\includegraphics[scale=.4]{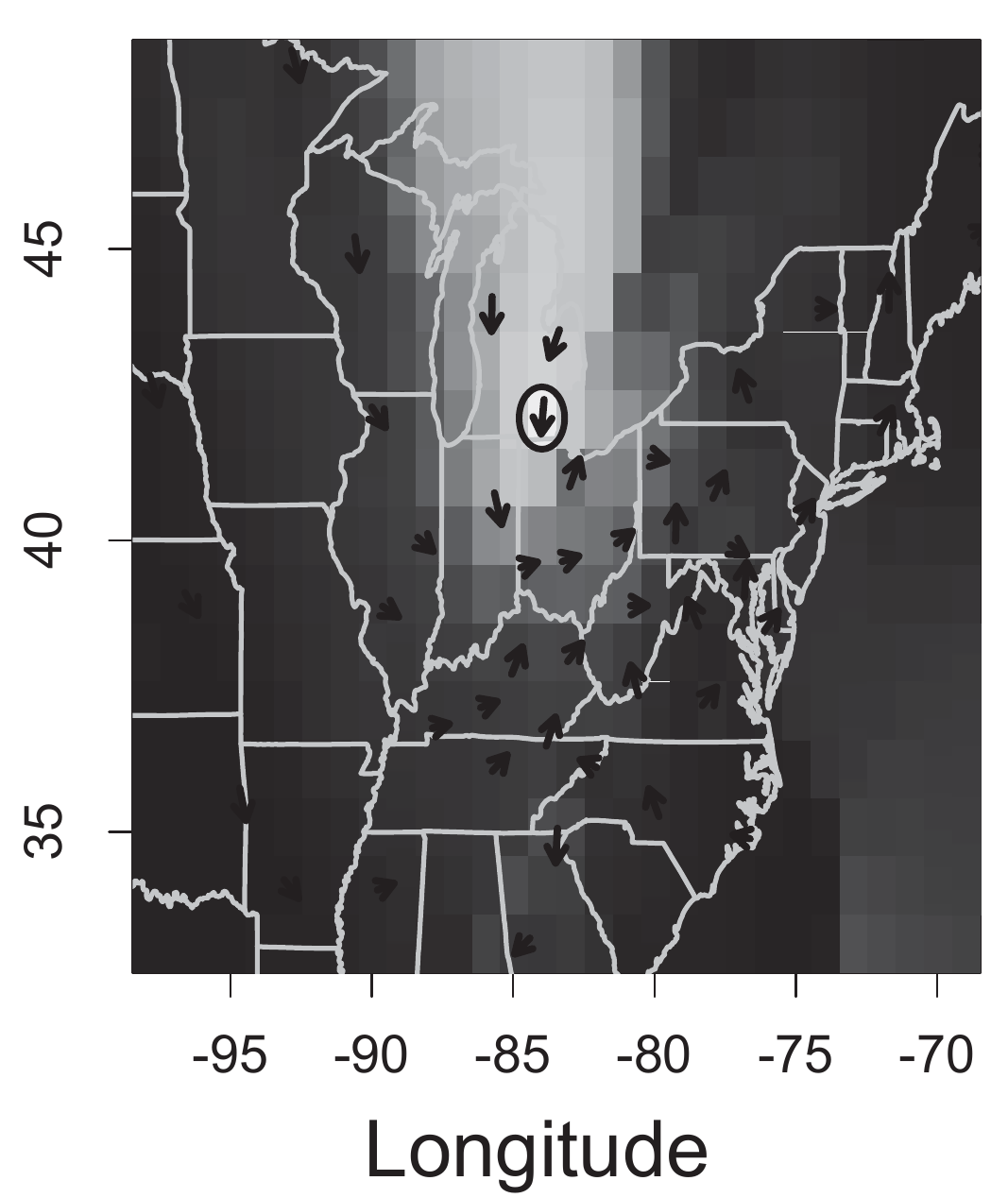}}
\subfigure{\includegraphics[scale=.4]{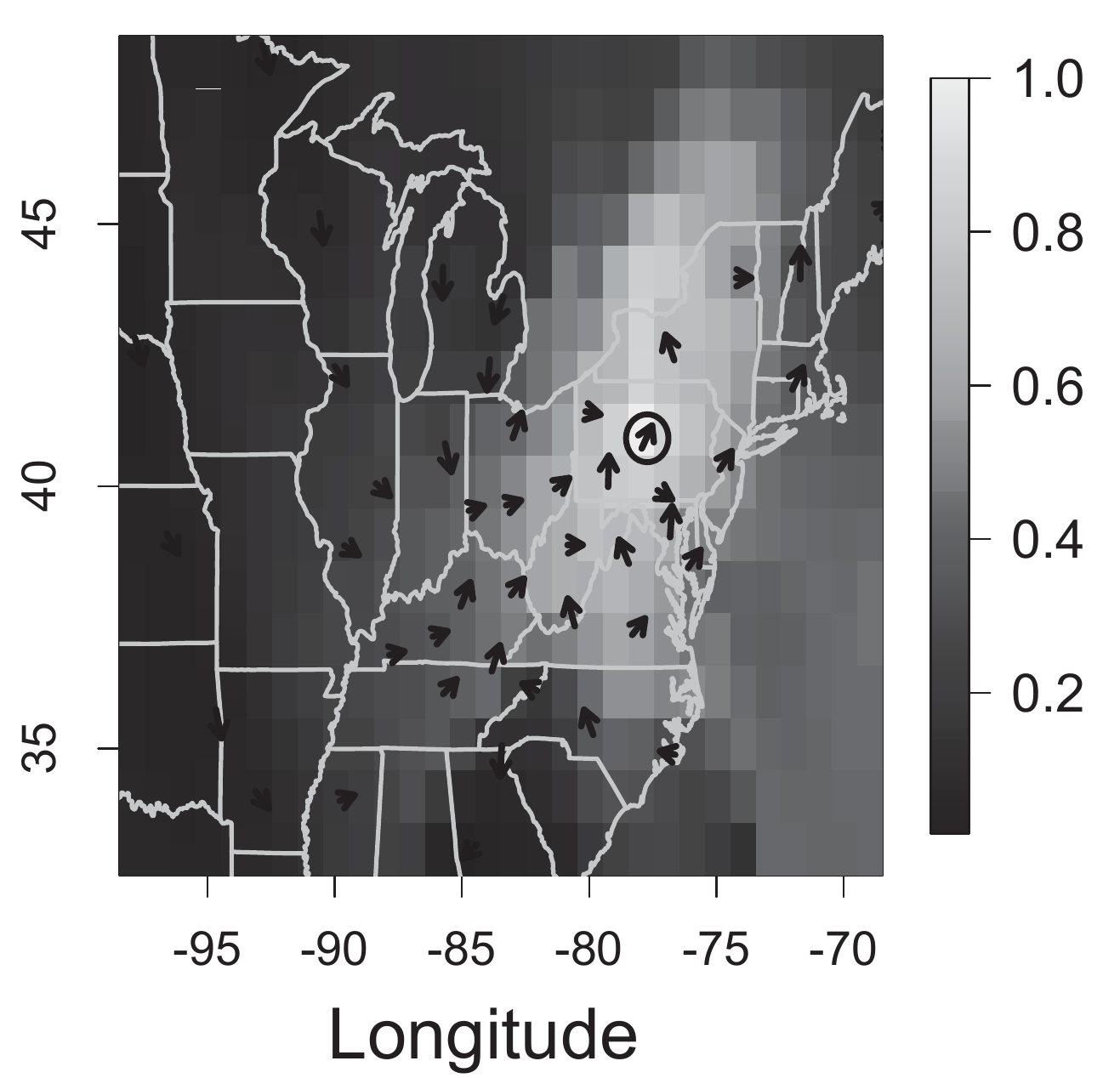}}
\end{center}
\caption{Posterior mean of the estimated correlation between the location marked by a circle and all the others in the grid, under models M3 (first row) and M4 (second row). Arrows indicate the direction at which the wind is blowing at the monitoring locations. The wind information at unmonitored locations was obtained through the spatial interpolation described in Section \ref{sec:predictive} and panel shown in Section C of the Supplementary Material. \label{fig:correlation}}
\end{figure}

The behavior of the proposed correlation functions in Sections \ref{sec:Gaussian} and \ref{sec:kernelsdiscretized} can be visualized through the panels of Figure \ref{fig:correlation}. The first row presents the estimated correlation for three different points, with all the others in the grid under model M3. And the panels of the second row show the estimated correlation for the same points under model M4. From these panels it is clear that the proposed models are able to capture the influence of the wind field on the estimated correlation function. 

\begin{figure}[!htb]
\begin{center}
\subfigure{\includegraphics[scale=.25]{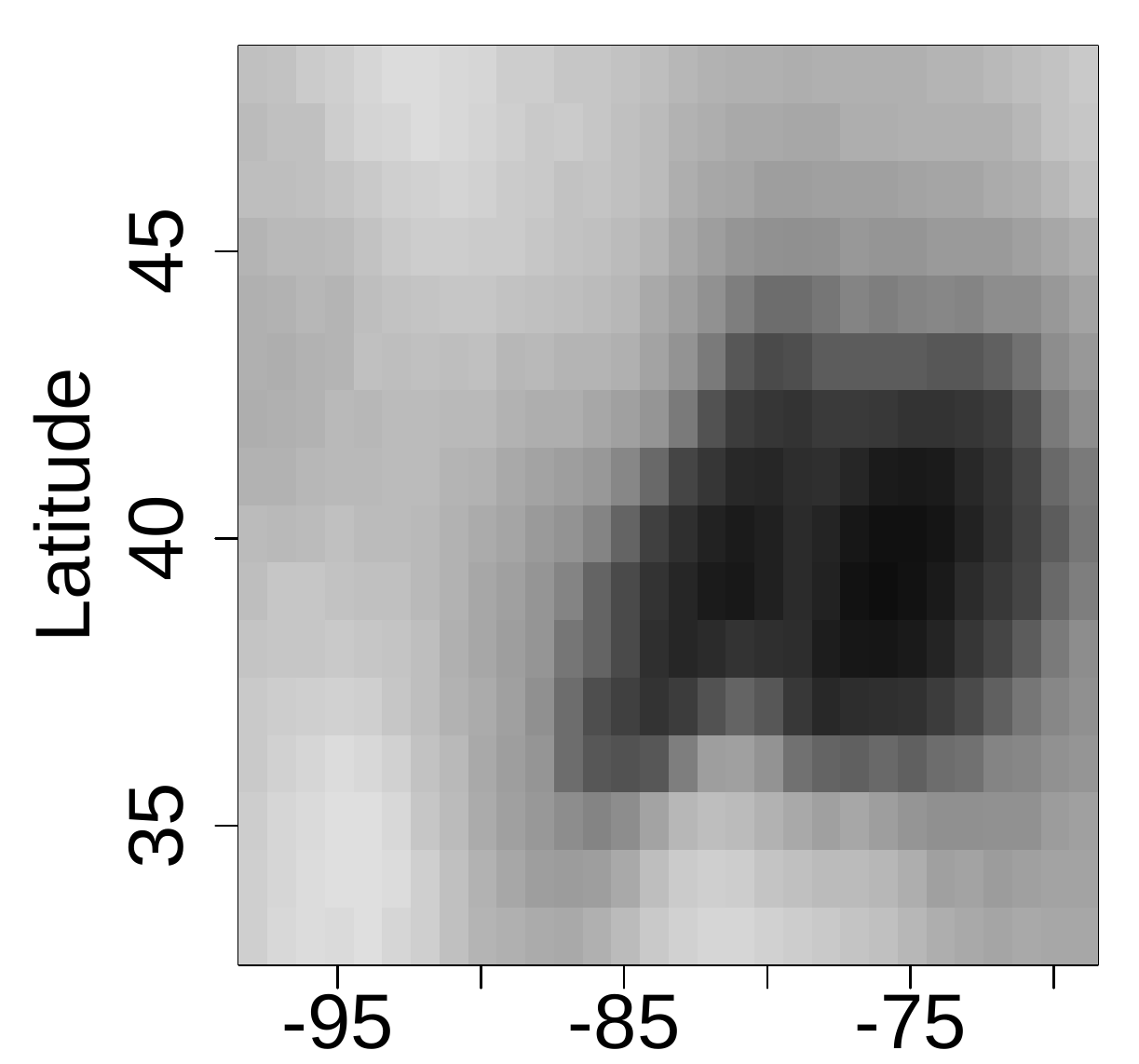}}
\subfigure{\includegraphics[scale=.25]{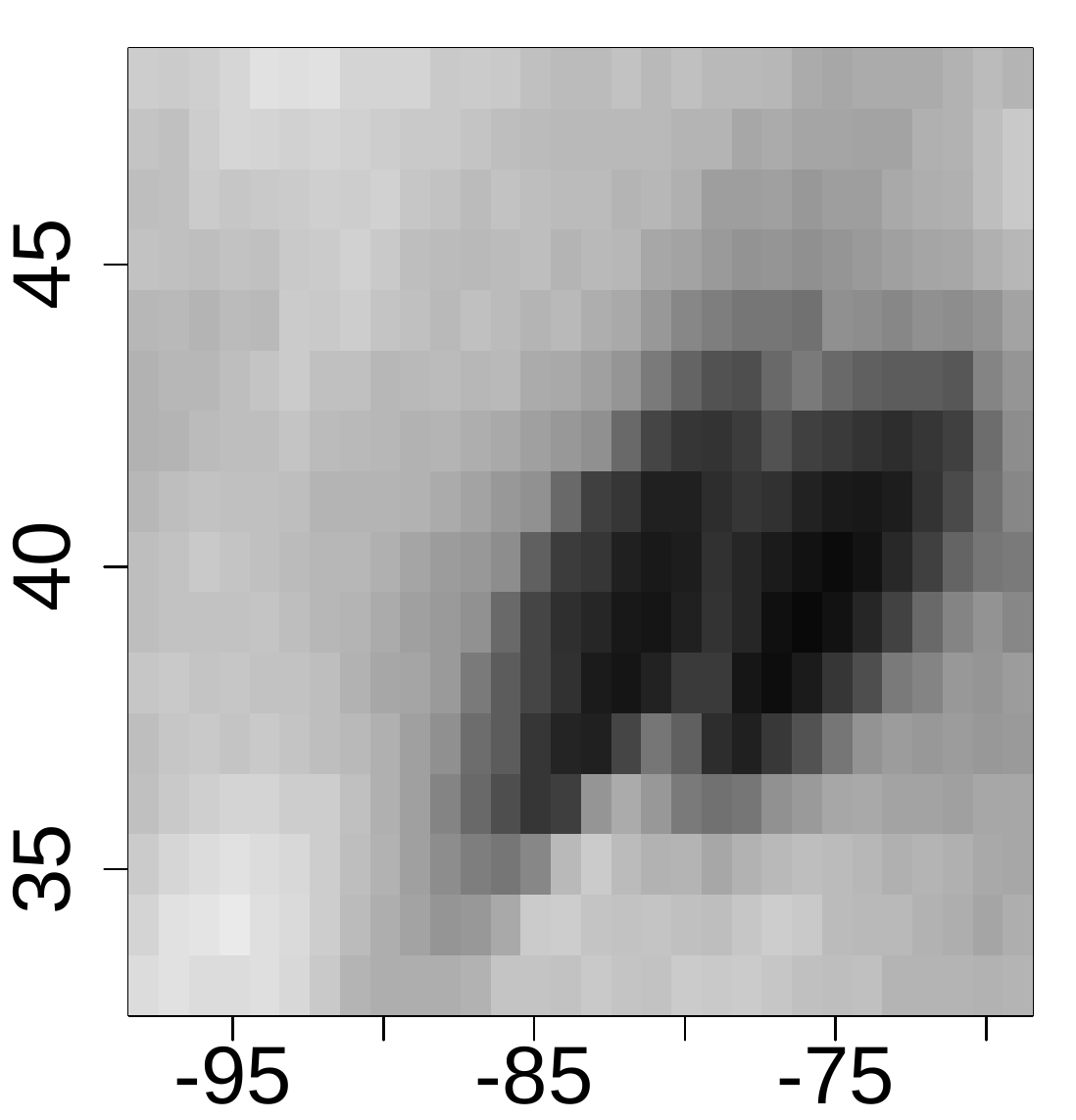}}
\subfigure{\includegraphics[scale=.25]{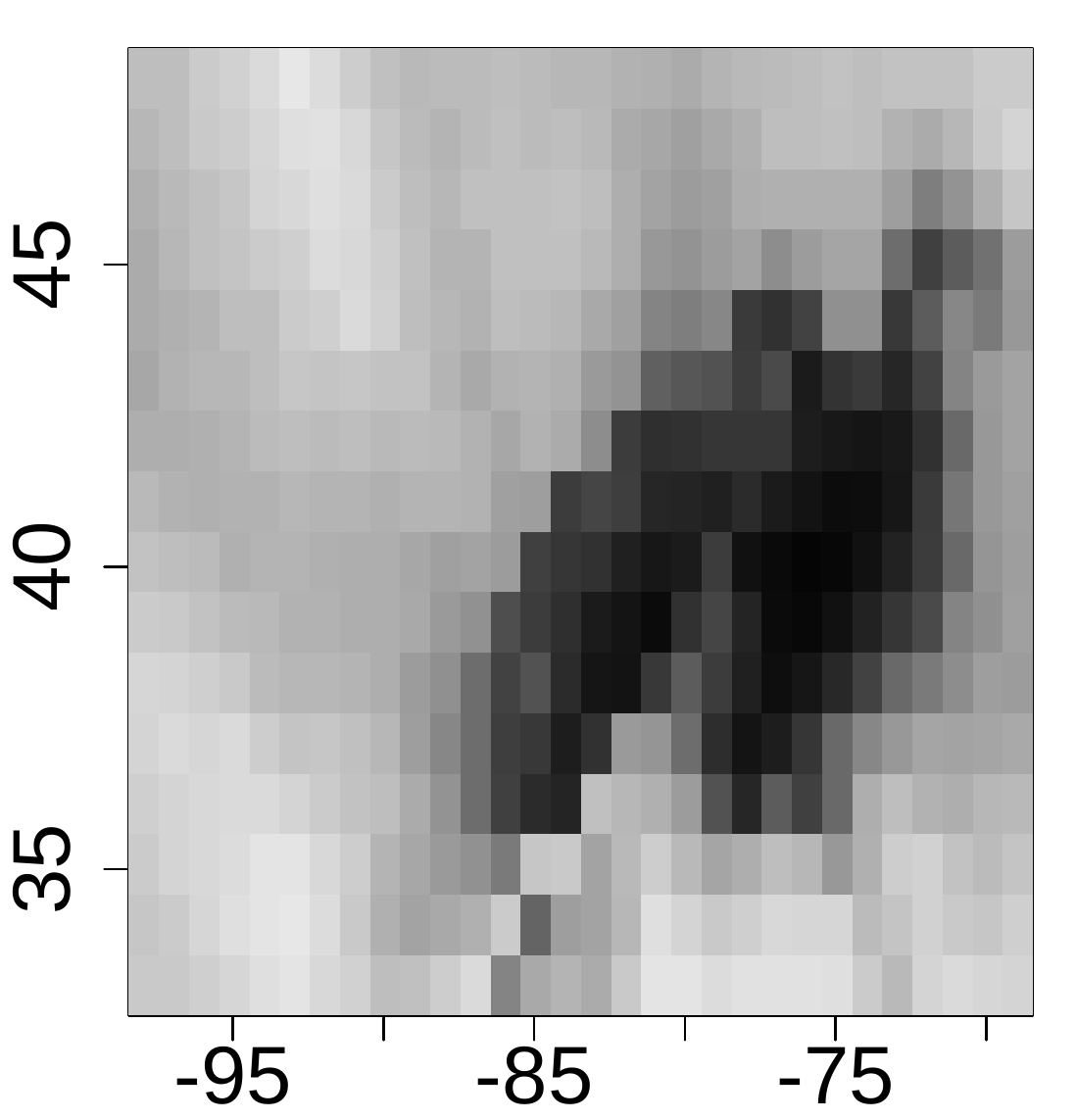}}
\subfigure{\includegraphics[scale=.25]{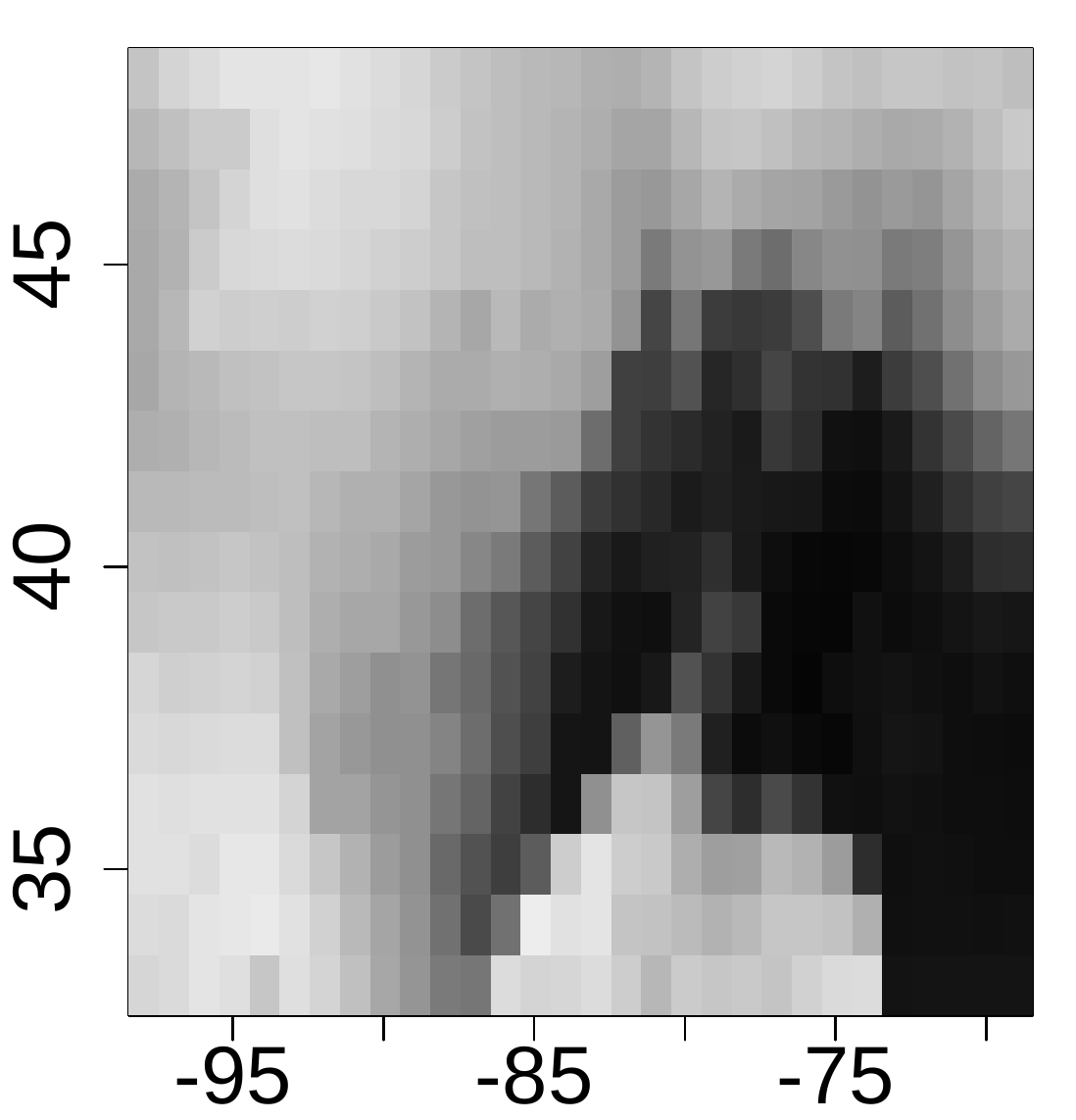}} 
\subfigure{\includegraphics[scale=.25]{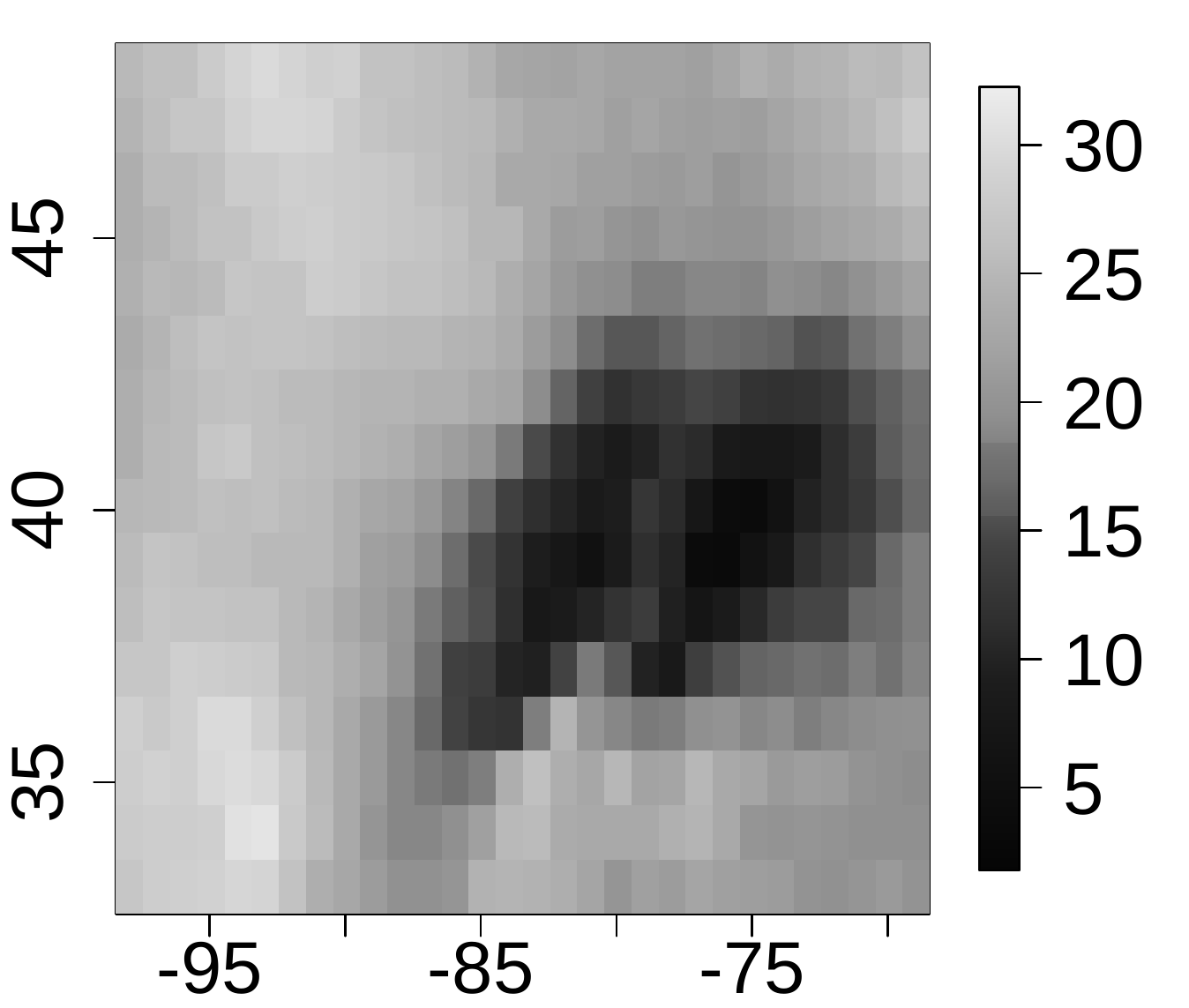}} \\
\setcounter{subfigure}{0}
\subfigure[M1]{\includegraphics[scale=.25]{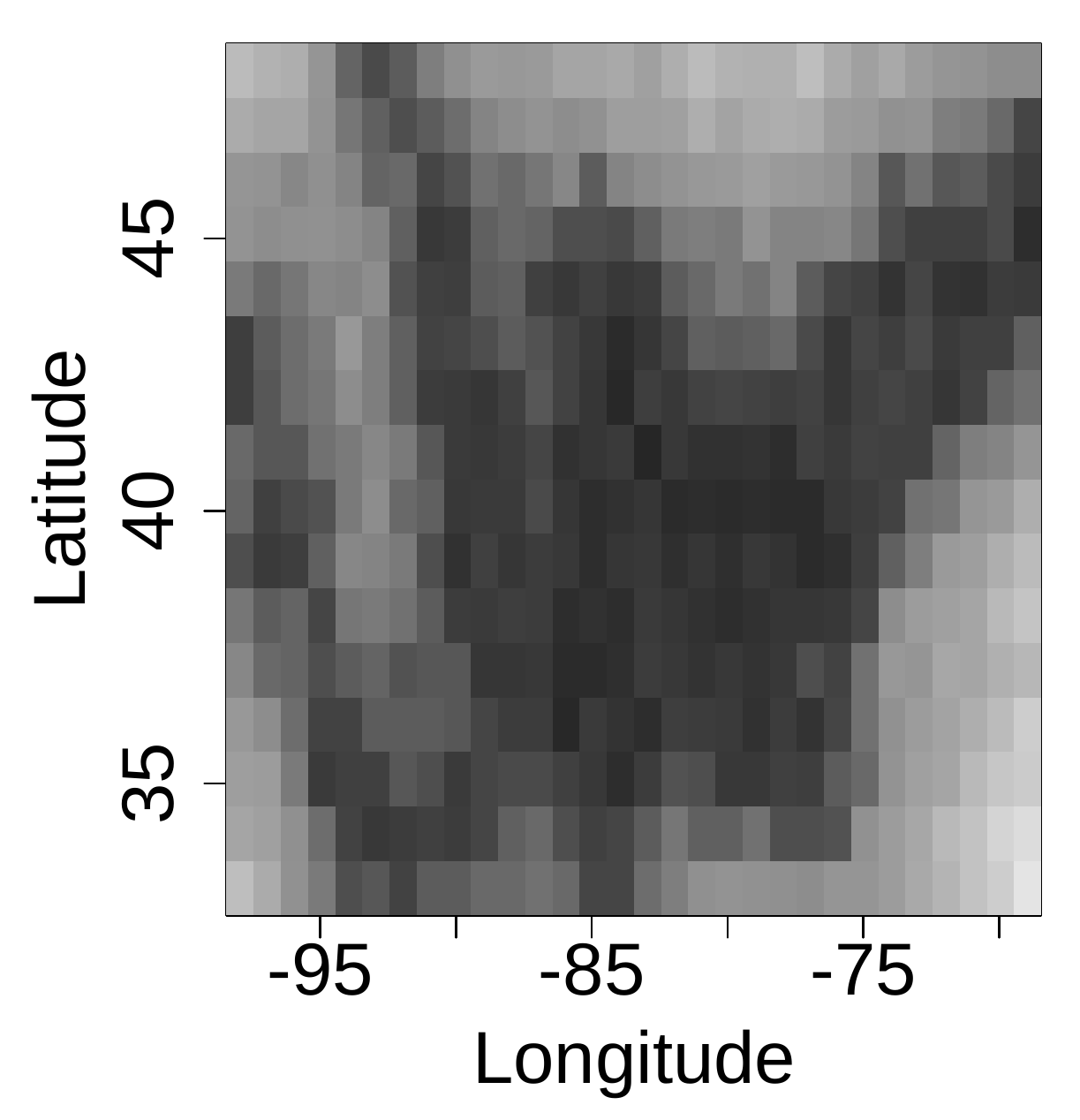}}
\subfigure[M2]{\includegraphics[scale=.25]{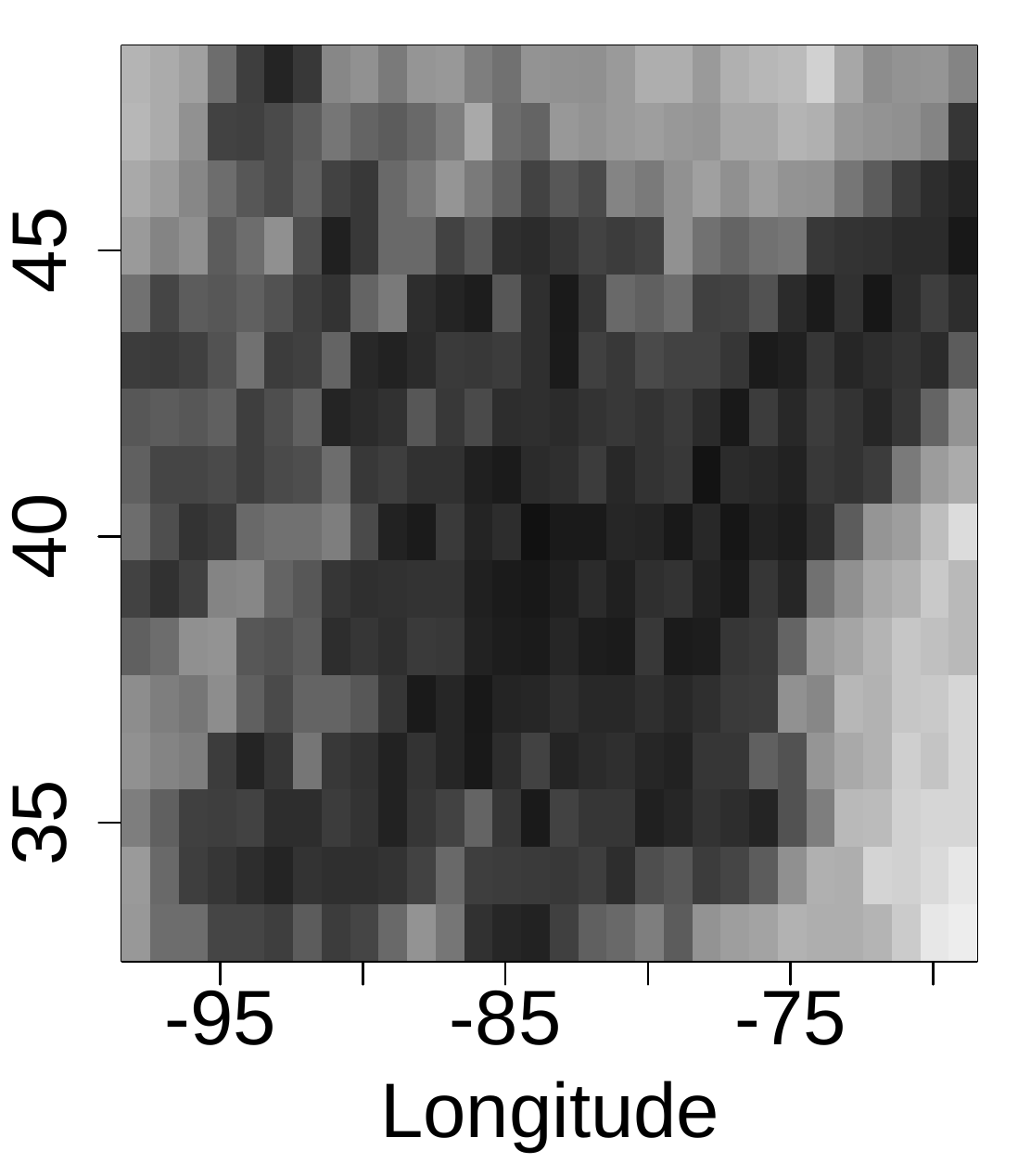}}
\subfigure[M3]{\includegraphics[scale=.25]{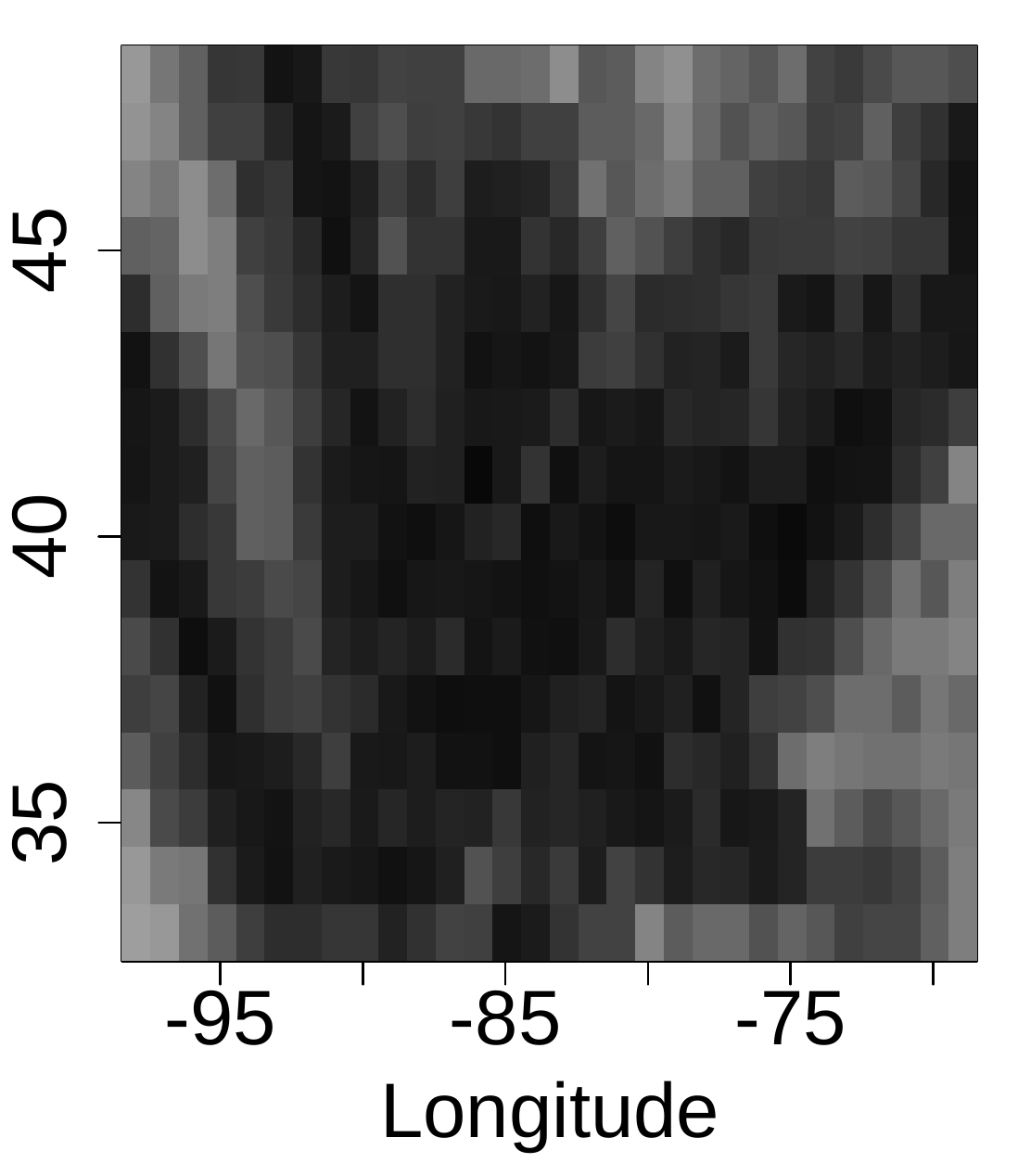}}
\subfigure[M4]{\includegraphics[scale=.25]{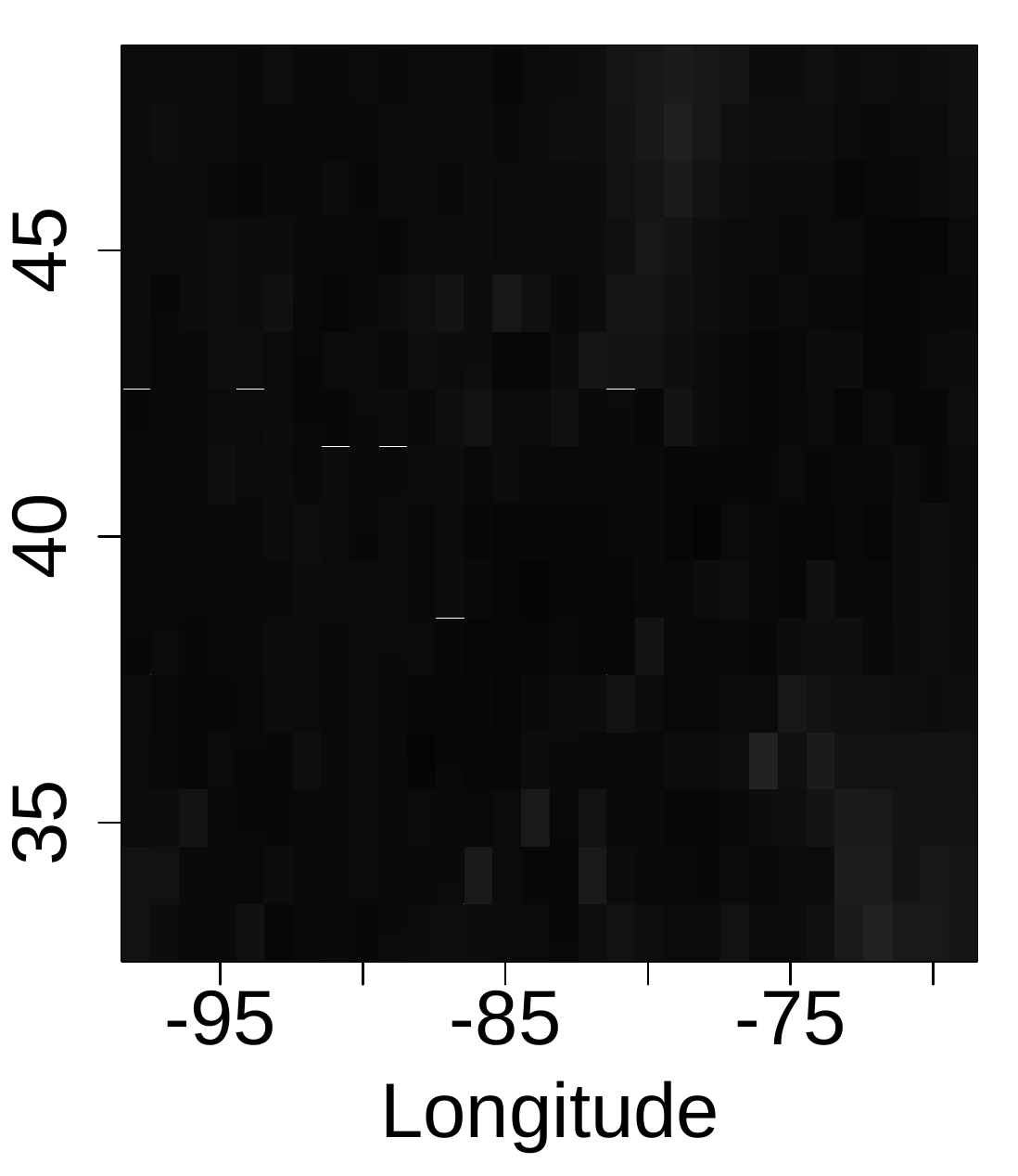}}
\subfigure[M5]{\includegraphics[scale=.25]{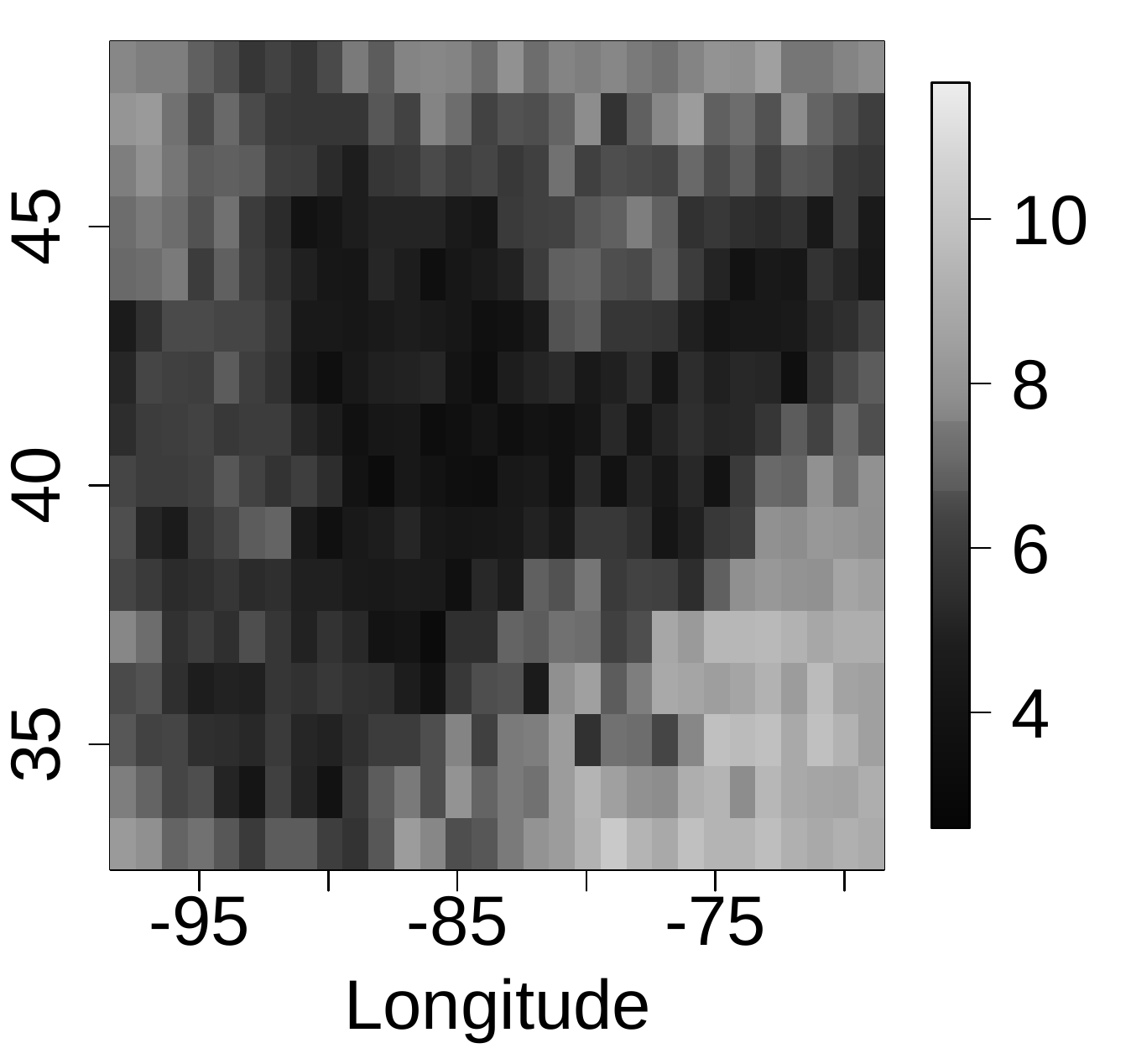}}
\end{center}
\caption{Surfaces of the posterior mean (first row) and standard deviations (second row) of the predictive distribution for unmonitored locations under each of the fitted models, M1-M5. \label{fig:surfaces}}
\end{figure}

Figure \ref{fig:surfaces} presents the  mean and standard deviation surfaces of the posterior predictive distribution of ozone for unmonitored locations formed by a regular grid superimposed over the region. For models M3 and M4, we interpolated the wind components, $U(.)$ and $V(.)$, based on the fitting of independent Gaussian processes as discussed in Section \ref{sec:predictive}.  The plot of the interpolated wind field can be seen in Figure 3 of the Supplementary Material. Now, the effect of the estimates of $\sigma^2$,  $\tau^2$, and the correlation structure of the spatial process $Y(.)$ under each fitted model is clearer. The spatial distribution of the standard deviations of the predictive distribution differ a lot across the models (second row of Figure \ref{fig:surfaces}). Models M1 and M2 tend to result in the biggest values of the standard deviations, while model M4 results in the smallest values, followed closely by model M3. The mean of the predictive distribution across the models are quite similar, with some differences in the southeast portion of the region.

\begin{figure}[!hbt]
\begin{center}
\subfigure[M1]{\includegraphics[scale=.28]{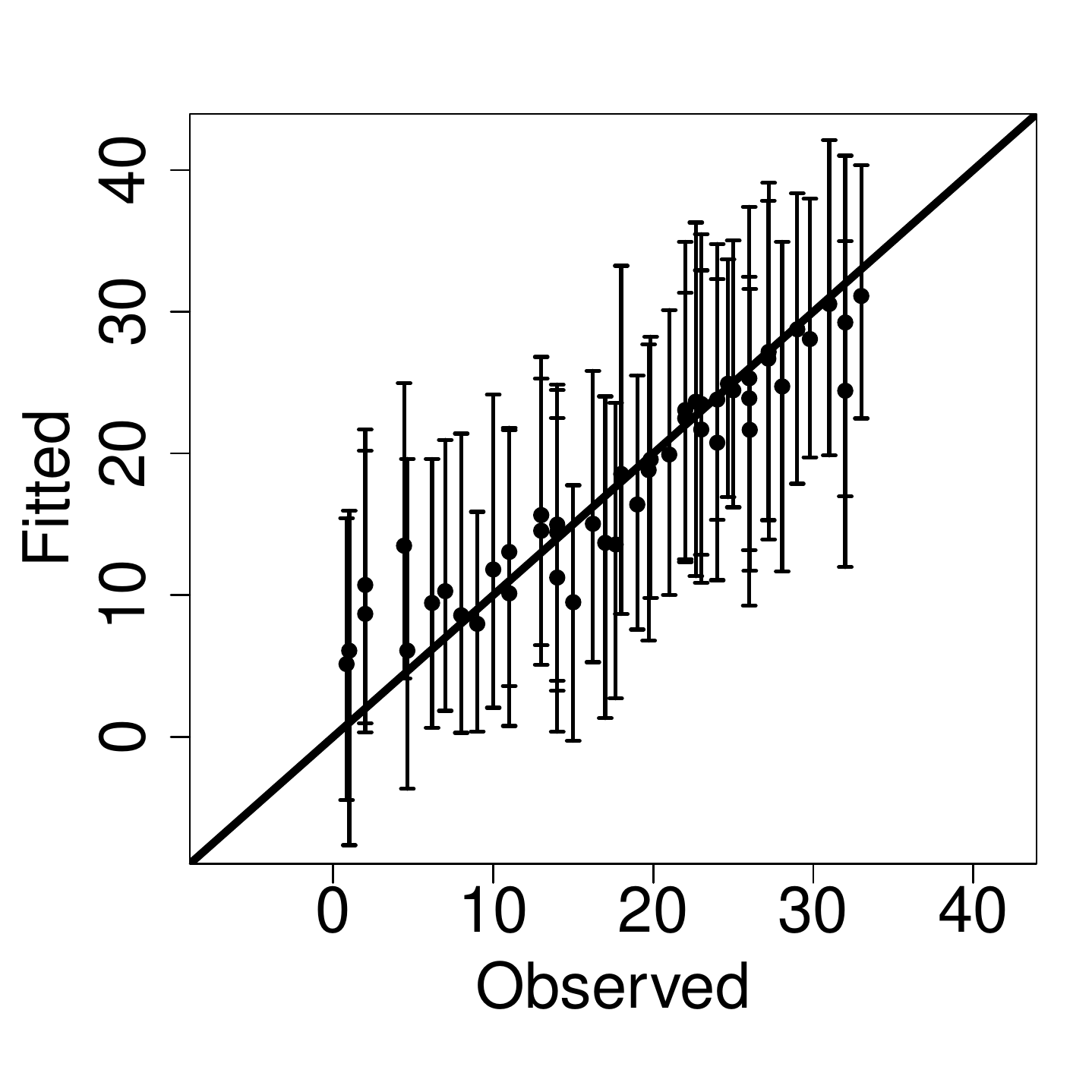}}
\subfigure[M2]{\includegraphics[scale=.28]{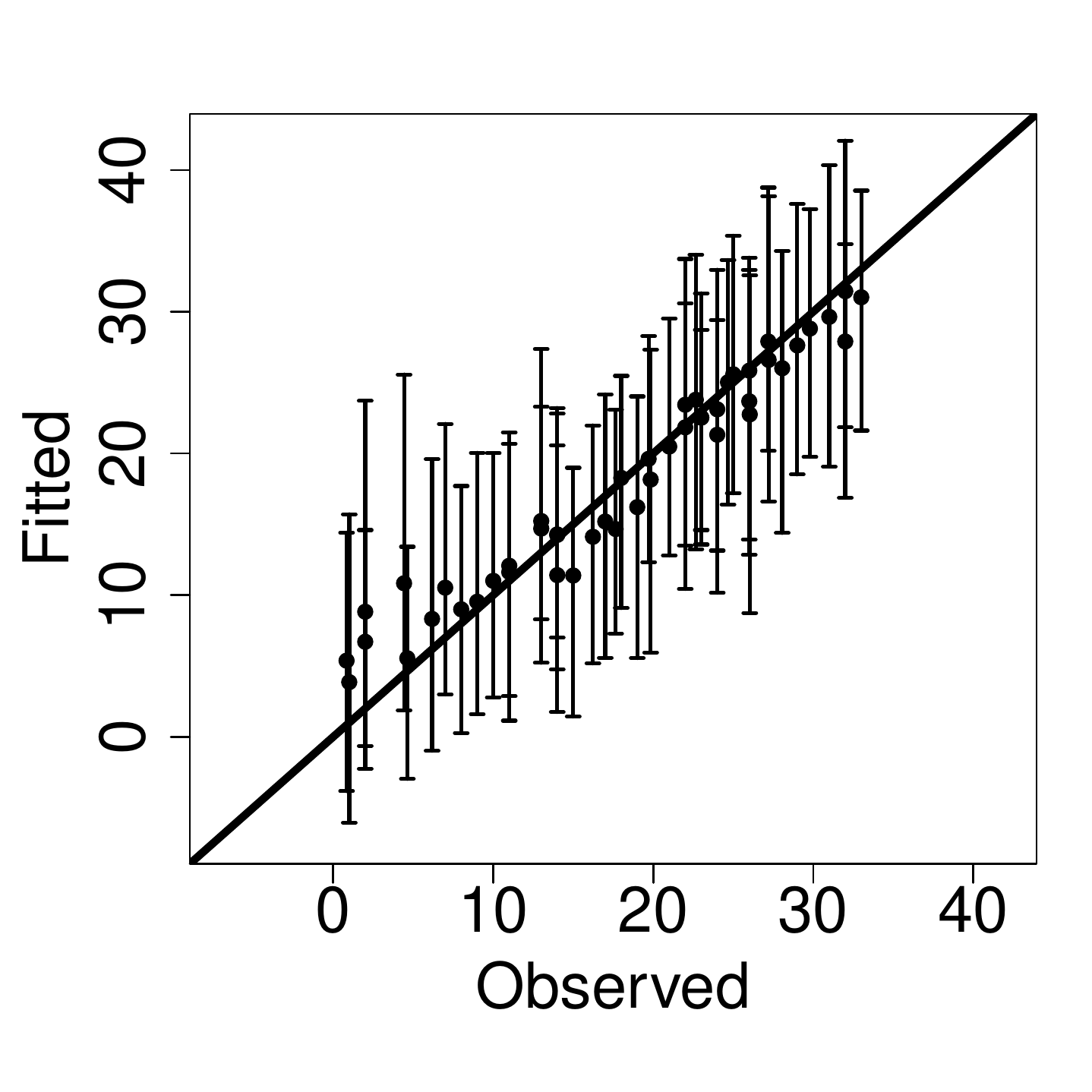}}
\subfigure[M3]{\includegraphics[scale=.28]{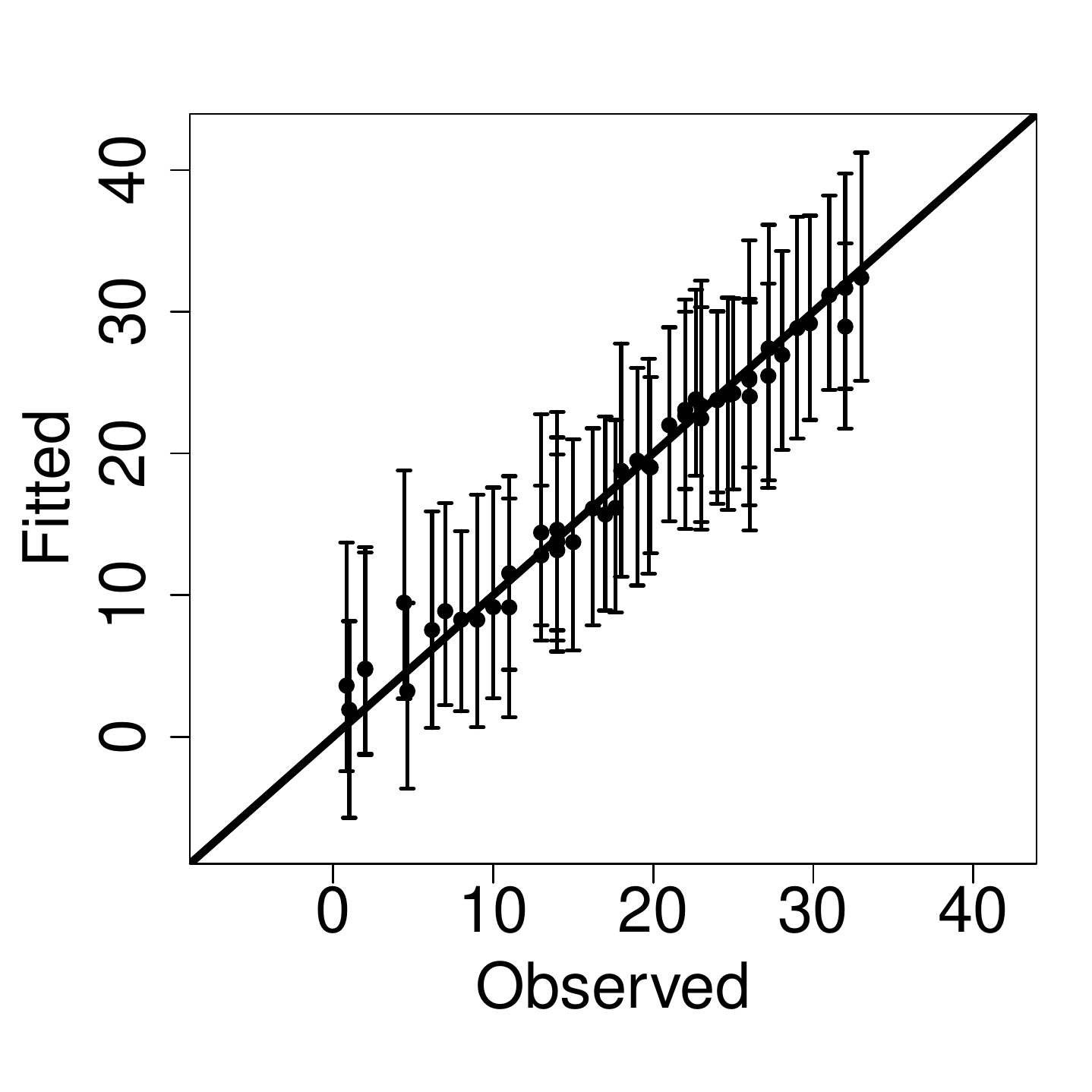}} \\
\subfigure[M4]{\includegraphics[scale=.28]{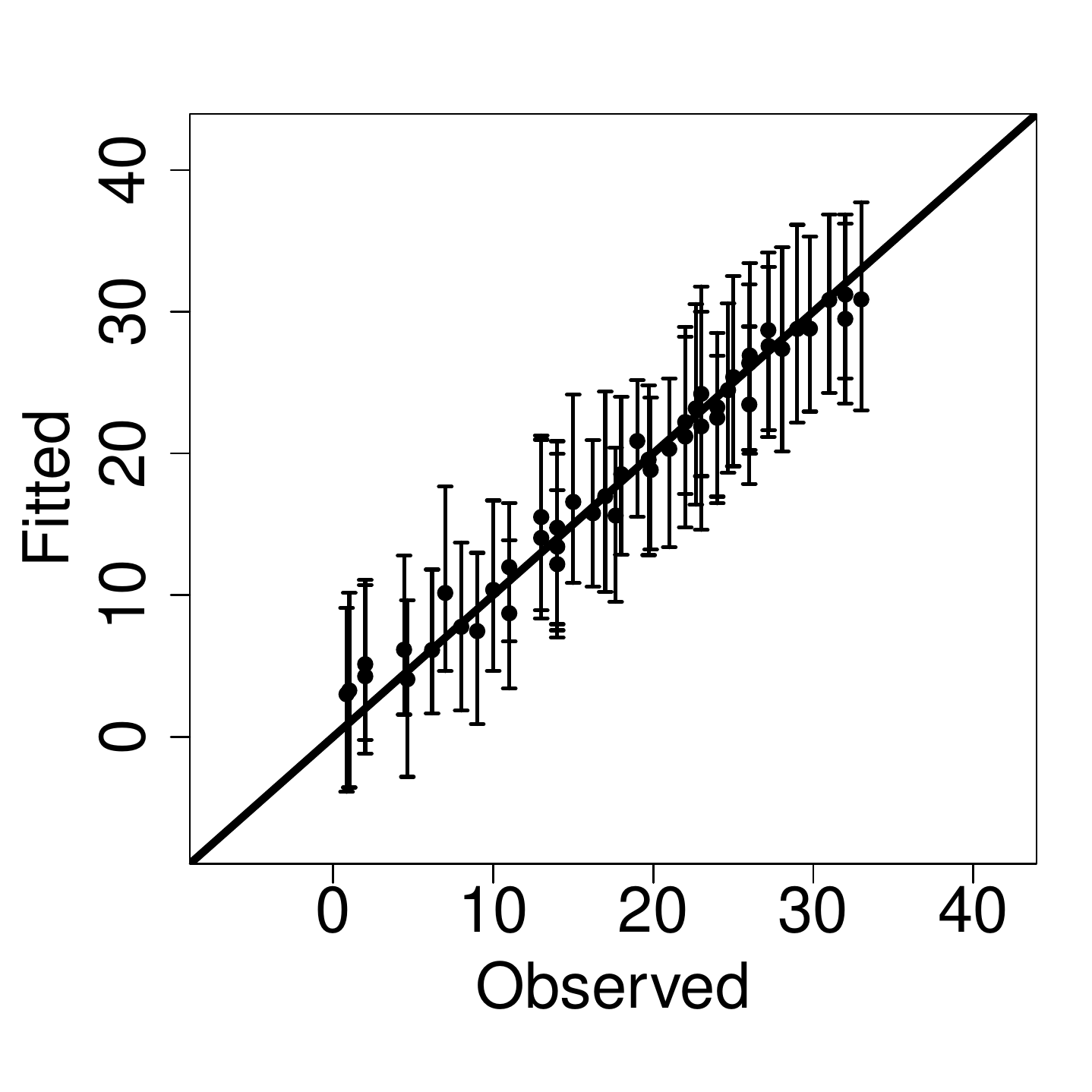}}
\subfigure[M5]{\includegraphics[scale=.28]{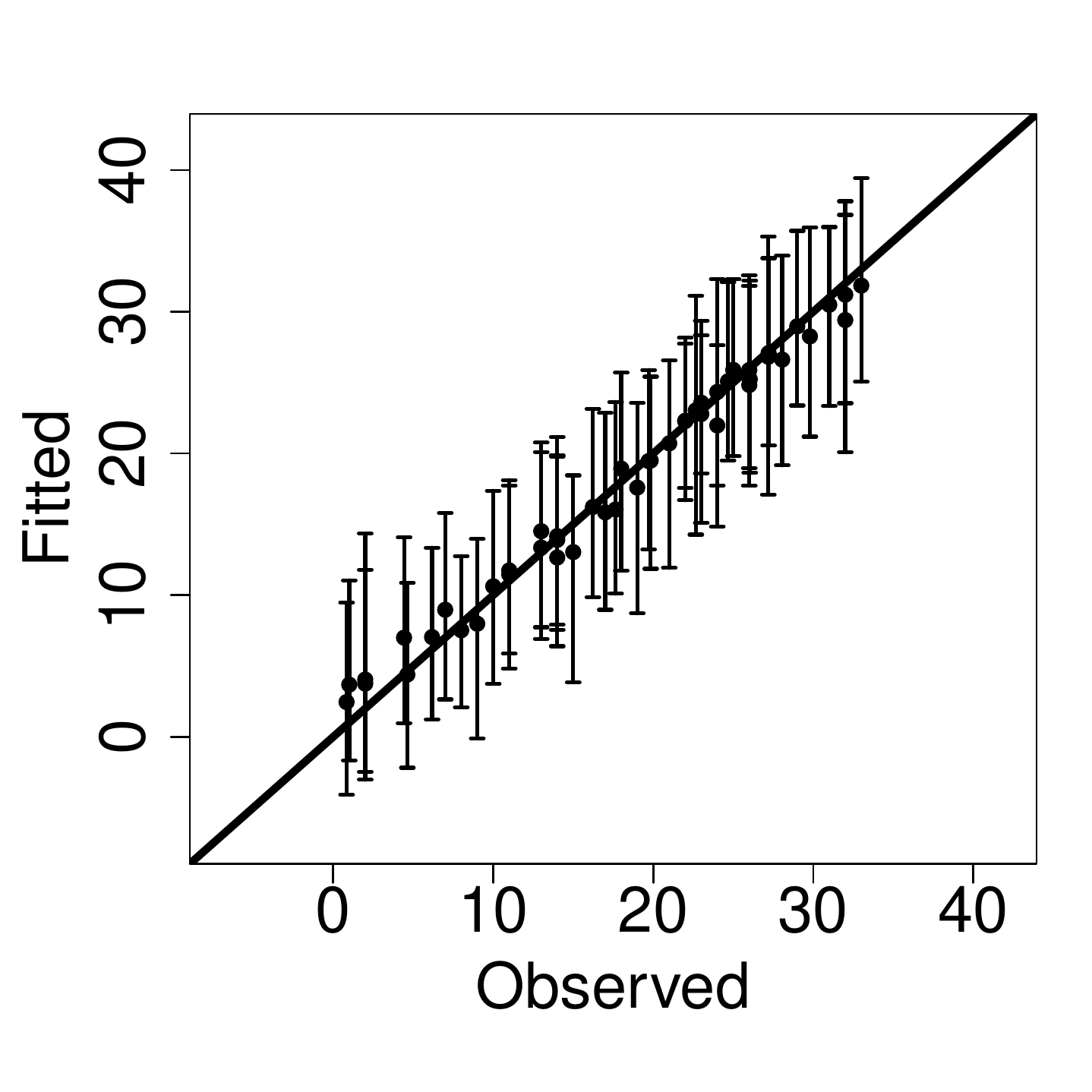}}
\end{center}
\caption{Posterior summary, mean (solid circle) and limits of the 95\% credible intervals (vertical line), of the observed values versus the fitted  ones, under each fitted models, M1-M5. \label{fig:obsxajus}}
\end{figure}

Finally, Figure \ref{fig:obsxajus} shows the posterior summary of the fitted values versus the observed ones. Clearly, despite of model M4 being much simpler than model M5, both tend to present very similar results in terms of model fitting, followed closely by model M3. From these panels the difference between the proposed models, M3, M4 and M5, and models M1 and M2 becomes clearer. Note that the ranges of the 95\% credible intervals under models M1 and M2 are much wider than those obtained under models M4, M5, and M3. This is another indication of the gain in considering more flexible covariance functions when modelling this  spatial process.

\clearpage

\section{Discussion}\label{sec:conclusao}

 Wind direction influences many environmental  processes of interest. We suggested two different approaches to consider wind information in the covariance structure of environmental processes observed at fixed locations. 
The challenge is how to account for the influence of the wind direction while still guaranteeing that the resultant spatial covariance structure is positive definite.
 Borrowing ideas from the convolution approach of \citeasnoun**{HIGDON1999}, and \citeasnoun**{PACIOREK2006}, we proposed to include the directional covariate in the kernel function of their convolution approach as discussed in section \ref{sec:proposta}.

 In particular, in Section \ref{sec:Gaussian} we described how to introduce information about wind  in the nonstationary Mat\'ern covariance function  proposed by \citeasnoun**{PACIOREK2006}. Although the resultant covariance function between two locations accounts for wind information at these  locations, it fails to account for the direction at which the winds are blowing at each point. To overcome this possible drawback, we proposed an alternative approach in Section  \ref{sec:kernelsdiscretized}. This proposal is based on an exponential kernel which depends on the Euclidean distance between locations and on the norm of a projection quantity which indicates how concordant two locations are in terms of the mean direction at which the wind is blowing at these locations. As the proposed kernel does not result in a closed form of the covariance function, we propose to use the discretized version of the convolution approach, as described in equation (\ref{eqn:discretized1}).

The use of the directional covariate significantly reduces the number of  parameters to be estimated, while still allowing for flexibile  covariance structures. In order to compare the performance of the proposed models in Sections \ref{sec:Gaussian} and \ref{sec:kernelsdiscretized},  Section \ref{sec:alternative} proposed  an alternative parameterization of the spectral decomposition of $\Sigma(s)$ in equation (\ref{eq:covarMatern}). Under this approach, different from \citeasnoun**{PACIOREK2006}, we  avoid the use of arbitrary constants, which are problem specific and difficult to fix when modelling  $\Sigma(s)$. When compared to the models of Sections \ref{sec:Gaussian} and \ref{sec:kernelsdiscretized}, this is a more flexible model, because it is able to capture different sources of nonstationarity. However, the great number of parameters to be estimated lead to some difficulties when obtaining samples from the resultant posterior distribution. More specifically, the chains of the parameters in $\Sigma(s)$ take very long to converge, requiring careful tunning of the variance of the proposal distributions in the Metropolis-Hastings step.

In Section \ref{sec:dataana} we fit five different models to ozone data observed at a particular time of a day in the East region of the USA. As expected, the MCMC algorithm to fit the model in Section  \ref{sec:alternative}   requires significantly  longer chains to provide evidence of convergence. For this particular dataset, the proposed models of Sections \ref{sec:Gaussian} and \ref{sec:kernelsdiscretized} perform better than standard models in the geostatistics literature. Moreover, although they have very few parameters to be estimated, they result in quite comparable fits to those obtained under the more flexible model of Section \ref{sec:alternative}, as shown in Table \ref{tab:comparison} and Figure \ref{fig:obsxajus}. 

The proposed model of Section \ref{sec:Gaussian}  requires the observation of wind at the same locations where the process of interest $Z(.)$ is observed. On the other hand, the model in Section  \ref{sec:kernelsdiscretized} requires the wind information at the points of the regular grid used to approximate the continuous process as proposed in equation (\ref{eqn:discretized1}). If the wind observations are not available at the locations of interest, one can perform  spatial interpolation of the wind field as suggested in Section \ref{sec:predictive}. It is common that wind information is available from satellite measurements which usually are available in the form of a grid of locations. When this is the case, we suggest to use this grid as the approximating grid of the model in Section \ref{sec:kernelsdiscretized}. 

The kernel functions proposed in Sections \ref{sec:Gaussian} and \ref{sec:kernelsdiscretized} can  be used in any other context for which  a directional covariate might influence a spatial process of interest. For example, when modelling a process observed at locations over an ocean, one can fit models with the kernel functions proposed here to investigate the influence of sea current on the spatial covariance function of such processes.

Introducing covariates in the covariance structure of spatial processes seem to provide reasonably flexible models, while significantly reducing the number of parameters to be estimated. From our experience, it is important to understand well the process of interest in order to propose which covariate, if any, might be used to obtain a flexible spatial covariance structure. In particular, when introducing  a directional covariate, such as the wind field, some care must be taken.
In order to visualize better what kind of fields, and correlations the covariance functions we propose here can provide, we made available, in the web link \url{http://www.ufjf.br/joaquim_neto/ensino/materiais/convolucoes-convolutions/},  R-TclTk softwares that produce graphical correlations,
covariances and simulations from the proposed models given values of the parameters specified by the user. We believe this tool might help the practitioner to check if the proposed models are able to reproduce what is expected to be observed in practice.

\section*{Supplementary Materials}

The reader is referred to the online Supplementary Materials for technical appendices.


\section*{Acknowledgements}

This work was part of the Ph.D. research of J. H. Vianna Neto under the
supervision of A. M. Schmidt. J. H. Vianna Neto was
partially supported by CAPES.  A. M. Schmidt was partially supported
by CNPq, grant no. 306160/2007-2. Schmidt and Guttorp are grateful
to {\em N\'ucleo de Apoio \`a Pesquisa em Modelagem Estoc\'astica e
Complexidade} (NUMEC)), USP, Brazil, for giving the opportunity to
discuss initial ideas on this project during the {\em Workshop on
Stochastic Processes Applied to Spatial Statistics: Multi-scenario
analysis and stochasticity in environmental prediction}. The authors thank two anonymous reviewers, an associate editor, and the editor whose comments and suggestions greatly 
improved the presentation of the paper.


  \bibliographystyle{dcu}
  \bibliography{Referencias}


  \end{document}